\documentclass{pas}

\usepackage{multirow}
\usepackage{amsmath}

\usepackage{amssymb}
\usepackage{comment}

\begin{document}

\lefttitle{Publications of the Astronomical Society of Australia}
\righttitle{Watters et al.}

\jnlPage{X}{X}
\jnlDoiYr{2026}
\doival{10.1017/pasa.xxxx.xx}

\articletitt{Research Paper}
\title{Critical Evaluation of Studies Alleging Evidence for Technosignatures in the POSS1-E Photographic Plates}

\author{\sn{Wesley Andr\'{e}s} \gn{Watters}$^{1}$, \sn{Laura} \gn{Domin\'e}$^{2,3}$,
\sn{Sarah} \gn{Little}$^4$, \sn{Cameron} \gn{Pratt}$^5$, Kevin H. Knuth $^6$, Matthew Szenher $^7$}

\affil{$^1$ Whitin Observatory, Dept. Physics and Astronomy, Wellesley College, 106 Central St., Wellesley, MA, USA 02482; ORCID 0009-0003-4052-3172}
\affil{$^2$ National Institute of Information and Communications Technology (NICT), Koganei, Tokyo, Japan; ORCID 0000-0002-2153-4728}
\affil{$^3$ Division of Physics and Astronomy, Graduate School of Science, Kyoto University, Kyoto, Japan}
\affil{$^4$ Scientific Coalition for UAP Studies, Fort Myers, FL, USA 33913}
\affil{$^5$ Scientific Coalition for UAP Studies, Fort Myers, FL, USA 33913; ORCID 0000-0002-6653-8490}
\affil{$^6$ Department of Physics, State University of New York at Albany; U. Albany Project X; ORCID 0000-0001-7702-3807}
\affil{$^7$ Scientific Coalition for UAP Studies, Fort Myers, FL, USA 33913; ORCID 0009-0008-2959-7994}

\corresp{W. A. Watters, Email: wwatters@wellesley.edu}

\citeauth{Watters W. A. et al., {\it PASA} {\bf XX}, X--X. https://doi.org/10.1017/pasa.xxxx.xx}

\history{(Received xx xx xxxx; revised xx xx xxxx; accepted xx xx xxxx)}

\begin{abstract}

Recent studies by B. Villarroel and colleagues have assembled and analysed datasets of unidentified features measured from digital scans of photographic plates captured by the first-epoch Palomar Observatory Sky Survey (POSS1) in the pre-Sputnik era. These studies have called attention to (i) a purported deficit of features within Earth's shadow; (ii) the sporadic presence of linear clusters; and (iii) a positive correlation between the timing of feature observations and nuclear tests as well as Unidentified Aerial Phenomena (UAP) sighting reports.  These observations were cited as evidence that some fraction of the unidentified features represent glinting artificial objects near Earth.  We have examined these claims using two previously published datasets that are closely related to those used in the Villarroel et al. studies. For these datasets, the Villarroel et al. assumption of a spatially uniform-random background distribution of features, essential to the Earth shadow analysis, is shown to be false.  After finding the null distribution of feature count deviations from the background, we find no statistically significant deficit in the shadow. We also determine that a third of the features in the reported linear clusters were not confidently distinguished from catalogue stars.  We find that the reported correlation between the timing of feature observations and nuclear tests becomes insignificant after properly normalizing by the relevant number of observation days, and is almost completely determined by the observation schedule of the Palomar telescope.  We uncover important inconsistencies in the definitions of the datasets used in these studies, as well as the use of unvalidated datasets containing catalogue stars, scan artefacts, and plate defects.  It has not been shown that any of the features in these datasets represent optical transients.  We examine the spatial distribution of the plate-derived features, finding an overall gradual increase in number density toward the corners and edges of plates, as well as examples of (i) empty north-south strips that span multiple plates; (ii) clusters and voids having geometric shapes; and (iii) amorphous clusters.  We also highlight a circular argument used in these studies, that leverages the results of an inferential analysis to justify conclusions about the origin of the features as well as the validity of the measurements. Finally, we also review the literature concerning historical searches for optical transients in photographic plates corresponding to gamma ray bursts (GRBs); following decades of work, researchers were unable to make a confident identification of a GRB-associated optical transient.
\end{abstract}

% UNCOMMENT for journal
\begin{keywords} optical transients, photographic plates, technosignatures, UAP, Unidentified Aerial Phenomena, Unidentified Anomalous Phenomena\end{keywords} \maketitle

\section{Introduction}\label{sec_intro}

In the past five years, the study of Unidentified Aerial Phenomena or Unidentified Anomalous Phenomena (UAP) has been acknowledged as an important scientific endeavour by several teams of academic scientists \citep{Kayal2022,nolan_etal-2021-001,loeb2023overview,watters2023scientific,szydagis2025initial} and by a NASA-commissioned report \citep{NASA2023}.  This recent work has built upon decades of prior scholarship and investigation by other academic scientists, government groups, and private research organizations \citep{ailleris2011lure,ailleris2024exploring,knuth2025new}.  It has been understood for decades that the vast majority of sighting reports (95-99\%) implicate known objects and phenomena.  A stubborn residuum has resisted conventional explanation \citep{hynek1966ufo,mcdonald-aaas1972,knuth2025new}.  This has inspired some scientists to investigate these phenomena using modern statistical methods, machine learning, and specially-designed scientific instrumentation, while suspending judgement about their ultimate origin \citep[e.g.,][]{teodorani-2004-001,watters2023scientific,domine2025commissioning,szydagis2025initial}.  In parallel, a subset of researchers engaged in the Search for Extraterrestrial Intelligence (SETI) has proposed searching for novel technosignatures within the Solar System \citep{haqq2012likelihood,shostak2020seti,villarroel2025cost}.  Many scientists participating in this new wave of research have stressed the importance of (i) imposing rigorous evidential standards, (ii) using calibrated instrumentation, (iii) investigating and completely understanding sources of error, and (iv) ensuring proper data validation \citep[e.g.,][]{watters2023scientific,lingam2023technosignatures,domine2025commissioning,szydagis2025initial}.  

\cite{villarroel2021exploring,villarroel2022glint} described searches in astronomical data records from the pre-Sputnik era for local technosignatures in the form of glinting artificial satellites. In the present study, we perform a critical examination of the data, analyses, and results presented in recent papers by \cite{villarroel2025aligned} and \cite{bruehl2025transients} that build on this previous work.  Both papers use datasets of nominally unidentified features that derive from datasets generated by \cite{solano2022discovering} using two sets of complete digital scans of the first-epoch Palomar Observatory Sky Survey (POSS1, 1949-1958): (i) the Space Telescope Science Institute's Digital Sky Survey\footnote{https://archive.stsci.edu/dss/acknowledging.html} (DSS) and (ii) the SuperCOSMOS survey \citep{hambly2001supercosmos}.  The first paper \citep{villarroel2025aligned} suggests that some fraction of the features may represent objects reflecting sunlight in near-Earth space. This conclusion is heavily based on an alleged deficit of features in the Earth's shadow. A set of candidate linear clusters is presented in this work; previous work was also concerned with the search for statistically significant alignments and clusters of the purportedly unidentified POSS1 features \citep{solano2024bright,villarroel2021exploring,villarroel2022glint}.  The second study \citep{bruehl2025transients} reports a correlation between the days when features were recorded and the timing of nuclear weapons tests as well as when UAP sightings were reported. We do not address UAP sightings in this paper.   

We shall use the term ``Selected POSS1-E features'' (SPFs) to refer to features selected from POSS1-E plates in these and other studies, where the selection criteria differed depending on the dataset in question.  We have done this partly in order to suspend judgement about the ultimate origin(s) of the features studied in \cite{villarroel2025aligned} and \cite{solano2022discovering}. In the present study, we examine three publicly available datasets of SPFs.
Two of these were catalogued in \cite{solano2022discovering}, and derive from the same parent dataset as the subsets analysed in \cite{bruehl2025transients} and \cite{villarroel2025aligned}.
For purposes of comparison, we also make use of a dataset of SPFs that appear in both the POSS1-E (red band) and POSS1-O (blue band) plates, which are hence overwhelmingly of celestial origin.  This third dataset was sampled as part of an independent study by the Minnesota Automated Plate Scanner (MAPS) project \citep{pennington1993automated,cabanela2003automated}.  We also use these publicly available datasets, along with information in \cite{bruehl2025transients} and \cite{villarroel2025aligned}, to describe the datasets addressed in those studies.  Access to the contents of these latter datasets is not necessary to derive the conclusions presented in this paper.

Regarding the SPFs used in \cite{villarroel2025aligned} and \cite{bruehl2025transients} to estimate correlations with the terrestrial shadow and nuclear tests, our work suggests that (i) a significant fraction (at least 91\%) of these belong to a set removed by \cite{solano2022discovering} because they reside within 5$^{\prime\prime}$ of catalogue stars; (ii) a potentially significant but undetermined amount are at risk of being star-like plate emulsion artefacts; and (iii) up to 40\% are at risk of being scan artefacts of the type removed by \cite{solano2022discovering} to obtain their most vetted dataset. \cite{villarroel2025aligned} 
acknowledged\footnote{e.g., See Section 8, pg 16 of \cite{villarroel2025aligned}: ``this sample... is expected to contain a substantial number of false positives, including clustered artefacts such as edge fingerprints or other plate defects that contaminate our sample.''} the possibility that a significant fraction of SPFs in the analysed datasets are the result of emulsion defects, degradation, and contamination, and did not formally validate these datasets in order to rule this out.

 Note that we have included some additional information in this study from a recent preprint \citep[][v2]{villarroel2026response} that provides new insights into the composition and spatial distribution of the primary dataset used in both \cite{bruehl2025transients} and \cite{villarroel2025aligned}. This information enhances the relevance of our analyses and confirms several of our findings; it is mentioned in various sections of our paper where applicable. We also take this opportunity to assure readers that the geometric correction cited in \cite{villarroel2026response} v2 was applied in all cases where this was required (see Section \ref{sec_boundary_dist}).

In Section \ref{sec_grbot_litreview}, we briefly review the literature about a two-decade search in archival plates for then-novel optical transients related to gamma ray bursts (GRBs), which revealed common star-like emulsion defects and the difficulties in  distinguishing these from optical flashes.  In Section \ref{sec_data_sources}, we summarise the principal datasets discussed in our study and provide detailed dataset definitions. In addition, we point out significant ambiguities and inconsistencies as well as caveats associated with their principal uses in \cite{villarroel2025aligned} and \cite{bruehl2025transients}, with Appendix \ref{app_data_sources} providing more details. 
Section \ref{sec_spatial_dist} addresses the spatial distribution of SPFs between plates, within plates, and across the sky, finding highly heterogeneous distributions (not Poisson-like), and suggesting plate, scan, and processing artefacts as their likely origin.  We also examine the SPFs reported to occur within sporadic linear clusters in \cite{villarroel2025aligned}, finding that a third of these SPFs were discarded by \cite{solano2022discovering}. Section \ref{sec_shaddow_freq} critically examines the deficit of SPFs in Earth's shadow reported in \cite{villarroel2025aligned} and offered as evidence that these represent glinting artificial satellites in the pre-Sputnik era. Using appropriate statistical analyses, we find no statistically significant deficit of features in the Earth's shadow, in either the most vetted or the largest datasets available from \cite{solano2022discovering}.
Section \ref{sec_nuclear} addresses the alleged correlation between the observation of SPFs (as a dichotomous variable) and the timing of nuclear tests as reported in \cite{bruehl2025transients}, finding that when the correct number of observation days is used, the significance drops, and that the residual correlation is almost entirely determined by the Palomar telescope observation schedule. 
The discussion in Section \ref{sec_discussion} focuses on major challenges to the analyses and conclusions in both of the studies we have examined, including the description of a circular argument used to justify confidence in the conclusions as well as the validity of measurements.

\section{Historical searches for optical transients in photographic plates}\label{sec_grbot_litreview}

\cite{villarroel2025aligned} and \cite{bruehl2025transients} analysed SPF datasets that we describe in Section \ref{sec_data_sources} and were motivated by questions developed in \cite{solano2022discovering}, \cite{solano2024bright}, \cite{villarroel2021exploring}, and \cite{villarroel2022glint}.  The goal of their work is to uncover evidence of technosignatures in the form of unidentified optical transients in pre-Sputnik photographic plates acquired by the National Geographic POSS1 survey \citep{minkowski1963national}.  The validity of this work rests on the premise that the selected features are in fact images of real optical transients (short-lived optical emissions) distinct from stars and other celestial bodies, and are not plate artefacts or other defects of spurious origin unrelated to objects in the sky.

Fifty years ago, a similar study sought to confidently identify optical transients in archival photographic plates \citep{grindlay1974search}. The goal of what became a two-decade-long effort was to identify the first optical flash counterpart of a gamma-ray burst (GRB). GRB optical counterparts were postulated as a new and unproven source of optical transients (OTs), similar to the role played by non-anthropogenic technosignatures in the work of Villarroel and colleagues, except that the GRBs were an independently verified phenomenon with estimates of position in the sky.  Work began in 1974 \citep{grindlay1974search}, but the first gamma-ray OT candidate was announced in 1981 after searching 4,135 physical plates from the Harvard University Observatory spanning 1889-1953 and 1970-1979 \citep{Schaefer1981probable}. Researchers then combed the physical plate collections of the Harvard, Sonneberg, Ondfejov, and Bamberg observatories ($\sim$910,000 plates; \citet{hudec1994optical}) and attempted to determine if any of the transient, star-like OT candidates found in the photographic emulsion of archival plates were: (i) the result of an optical flash; (ii) not from any known atmospheric or celestial event; and (iii) a counterpart to a GRB at a previously known location (see for example: \citet{greiner1987search, zytkow1990there, schaefer1990gamma, greiner1992rejection, schaefer1994search, greiner1994re, hudec1993optical, vrba1995deep, fishman1995gamma}). 

Researchers understood that a valid candidate for a GRB optical counterpart from an archival plate must first be established to be the result of an optical flash and not (i) a plate fault: crack, dent, scratch, static discharge halo, glass corrosion, emulsion backing fragment, endemic emulsion defect (spontaneous cluster of various sizes, shapes, and structures of silver grains \citep{schaefer1990gamma} due to manufacturing, exposure, or developing) that created a star-like false image \citep{hudec1993optical, varady1992background, schaefer1990gamma, greiner1990discrimination, greiner1987search}, or ageing blemish \citep{greiner1990discrimination}; or (ii) an instrument or processing artefact: double exposure, pointing instability, or instrument vibration \citep{hudec1993optical}. 

Second, these researchers needed to establish that the candidate was not due to any known real object: star \citep{schaefer1990gamma}, nova or dwarf nova \citep{hudec1994optical}, head-on meteor, satellite flash, aircraft flash \citep{hudec1993optical, varady1992background}, stellar flare, planetoid/minor planet, atmospheric event \citep{hudec1993optical}, or insect (e.g., firefly) \citep{schaefer1987perseus}.

Initial assessments about the validity of each star-like OT candidate centred on techniques that helped distinguish between a star-like plate artefact and a true optical transient. Ideally, a blink test was performed (which finds differences between two or more simultaneous photographs containing the same candidate). Without this, it was necessary to conduct an analysis of the physical plate itself using microscopy (500$\times$) and microdensitometry (to obtain photodensity image profiles) with transmitted and reflected light (to look at grain structure in the emulsion), as well as CCD scanning techniques. These techniques were used in an effort to find: (i) a valid star-like 3-D structure in developed silver grains within the plate emulsion, and evidence of short duration  (evidence of an optical flash); (ii) coma distortion (evidence of optical path); (iii) halos from atmospheric fog (evidence of optical path) \citep{hudec1994optical}; and (iv) lack of a trailing (caused by telescope tracking errors), which might indicate a short duration optical flash, but could also be a plate defect \citep{schaefer1990gamma}.

A number of real optical flashes were found associated with known phenomena.  For example, one GRB study of 32,000 plates \cite{schaefer1990gamma} found four OTs associated with variable sources in their control search area.

In the absence of multiple simultaneous plates, GRB OT candidates with the supplementary support provided by validation techniques were more likely to be accepted by scientific peers \citep{hudec1993optical}. However, different groups applied these techniques to the same plate and arrived at different conclusions \citep{hudec1993optical, vrba1996searches}. By 1995, several OT candidates were still being championed as real optical flashes, although not unequivocally associated with GRBs \citep{vrba1996searches}. 

Although nearly all plate defects could be rejected upon close inspection, a select few star-like candidates could not. Twenty years of dedicated searching for optical flashes near GRBs on physical, archival plates yielded five significant OT candidate sets (representing ten OTs). Among these, however, none could be ``either definitely proved or definitely excluded'' \citep[][pg 367]{hudec1993optical} as real optical transients representing a new astronomical phenomenon, and none could be convincingly shown to be a GRB counterpart \citep{vrba1996searches}. The researchers were unable to rule out all of the alternate hypotheses, and thus could not disprove the null hypothesis in their search for the first example of a GRB optical counterpart.

In 1996, a dedicated satellite was launched to rapidly and accurately locate GRBs \citep{mcnamara1998optical}, enabling potential OT counterparts to be imaged with CCD cameras in real-time and alleviating the need for the GRB community to work from the archival plates. 

The POSS1-E physical plate set of SuperCosmos copy negatives is housed at the Royal Observatory, Edinburgh \citep{hambly2024nature}. For many surveys, observations of star fields are repeated if there are quality concerns (e.g. inferior seeing), a practice that generates slightly inferior original plates that can be made available for detailed examination \citep{morgan1995sky}.

The DSS digitization of POSS1-E and -O plates began in 1992 and they became available on CD-ROM in 1995; the SuperCosmos digitization of POSS1-E plates became available in 2000. These DSS and SuperCosmos digital scans used in \cite{solano2022discovering} and \cite{villarroel2025aligned} added additional pathways for processing-related artefacts to enter the picture, such as images of dust particles and emulsion backing fragments. 

The challenging results from two decades of GRB OT studies, and the very few optical transient candidates found, suggest that a search for alignments and correlations of optical flashes from non-anthropogenic technosignatures should include physical plate-level validation, independent of the technosignature hypothesis, to ensure that the detections in the dataset(s) could plausibly represent optical flashes. 

Expressing similar concerns, \cite{hambly2024nature} examined the intensity profile statistics (position, brightness, and morphology) of all objects on the POSS1-E plate E0070 containing the nine transients allegedly identified by \cite{villarroel2021exploring}. This study found a class of objects they designated as ``spurious,'' that are statistically distinctive, and morphologically distinguishable from stars and other astronomical objects; e.g., they exhibit a smaller full-width at half maximum (FWHM) on average. They then trained a machine learning model to distinguish between stars, galaxies, and likely spurious (not within $5^{\prime\prime}$ of a Gaia or Pan-STARRS DR1 catalogue object) and applied it to plate E0070, finding $\sim 45,000$ stars, $\sim 27,000$ galaxies, and $\sim 8000$ spurious objects (including the nine ``transients'' from \cite{villarroel2025aligned}). Their so-called spurious objects nominally belong to the same class as the so-called transient dataset (N=298,165) defined by \cite{solano2022discovering} and used in \cite{villarroel2025aligned}. \cite{villarroel2025image} has since acknowledged that many of the unidentified SPFs exhibit a smaller FWHM when compared with stars, but have countered that this is not unexpected for short-lived flashes from objects in high-altitude orbits, during a long-duration ($\sim 45$ min) exposure.  The following critical questions remain unanswered after \cite{hambly2024nature}: (i) whether optical flashes exhibit smaller FWHM on POSS1-E plates; (ii) whether any of the 8,000 features selected on plate E0070 by \cite{hambly2024nature} represent an optical flash; and (iii) whether some categories of plate artefacts exhibit a smaller FWHM when compared to stars, on average.

\section{Data sources}\label{sec_data_sources}
In this section, we describe (i) the publicly available datasets used for the analyses in this paper; (ii) the publicly known properties of the related and as-yet unpublished datasets used by \cite{bruehl2025transients} and \cite{villarroel2025aligned}; and (iii) initial data quality concerns.  
Access to the unpublished datasets is not necessary for our evaluation of the analyses and claims in these papers. We use our own naming convention when referring to the datasets, using the symbols $A, S, W, P, V$, and $R$ (column 1 in Table \ref{tab_data_sets}), which are not used in the papers we have cited.

\begin{figure}[h]
\begin{center}
\noindent\includegraphics[width=20pc]{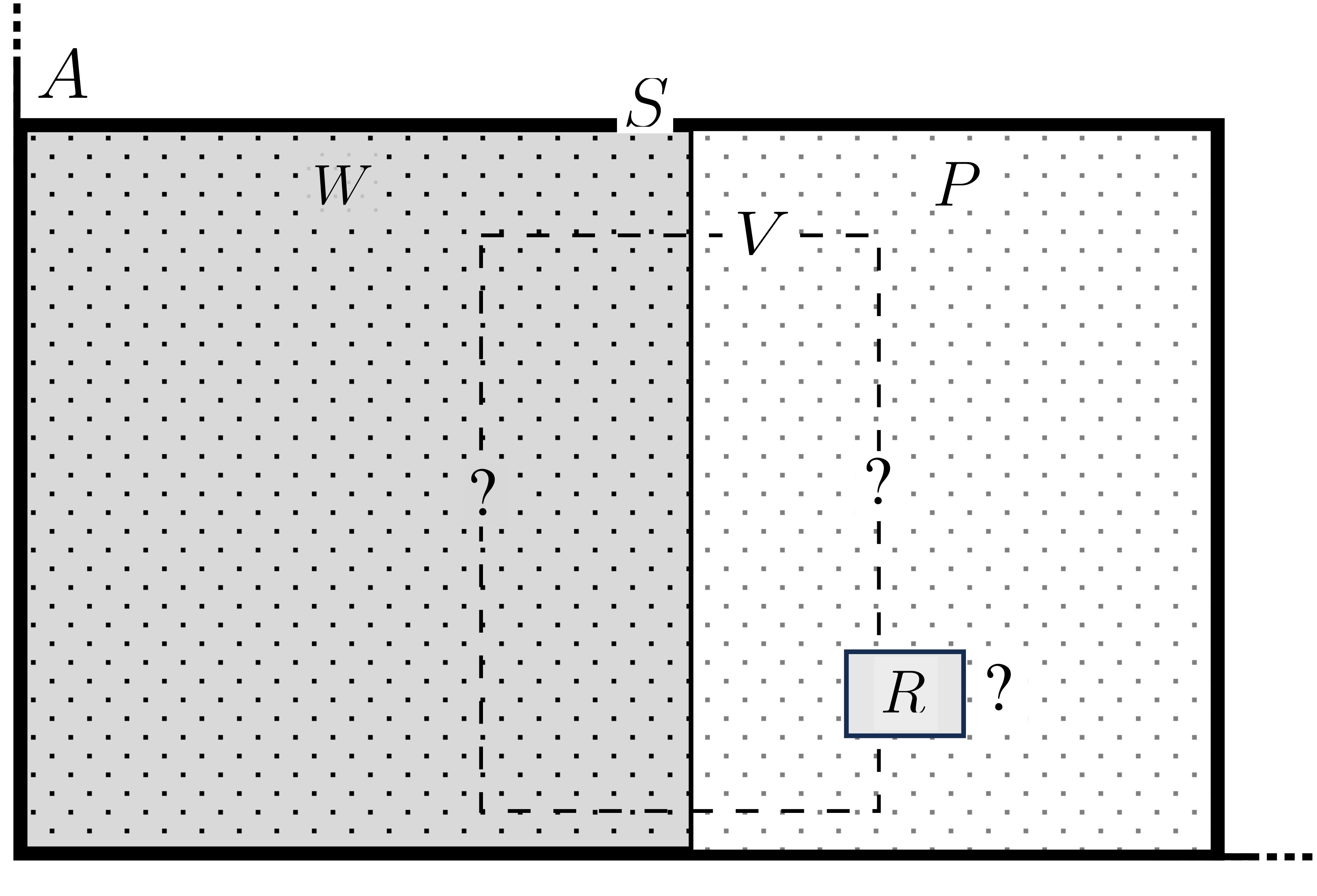}
\caption{Diagram illustrating the relative sizes and overlap relationships of the SPF datasets defined in Table \ref{tab_data_sets}.  The dashed line and ``?'' here represent our uncertainty about the degree of overlap between $V$ (the main focus of calculations in \cite{villarroel2025aligned} and \cite{bruehl2025transients}) and $W$ as well as $R$ (the vetted ``remainder'' of unidentified objects, defined in \cite{solano2022discovering}).  The shaded subsets $R$ and $W$ are publicly available.}
\label{fig_dbvenn}
\end{center}
\end{figure}

The relationships between these published and unpublished datasets are illustrated in Figure \ref{fig_dbvenn}, and defined in Table \ref{tab_data_sets}, which is structured to follow, when possible, the filter pipeline described by \cite{solano2022discovering}. Due to limited space, the following  citation abbreviations are used in the table: \cite{solano2022discovering} (SVR22); \cite{villarroel2025aligned} (VEA25); \cite{villarroel2026response} v2 (VEA26); and \cite{bruehl2025transients} (BV25).

\begin{table*}
\noindent\includegraphics[width=42pc]{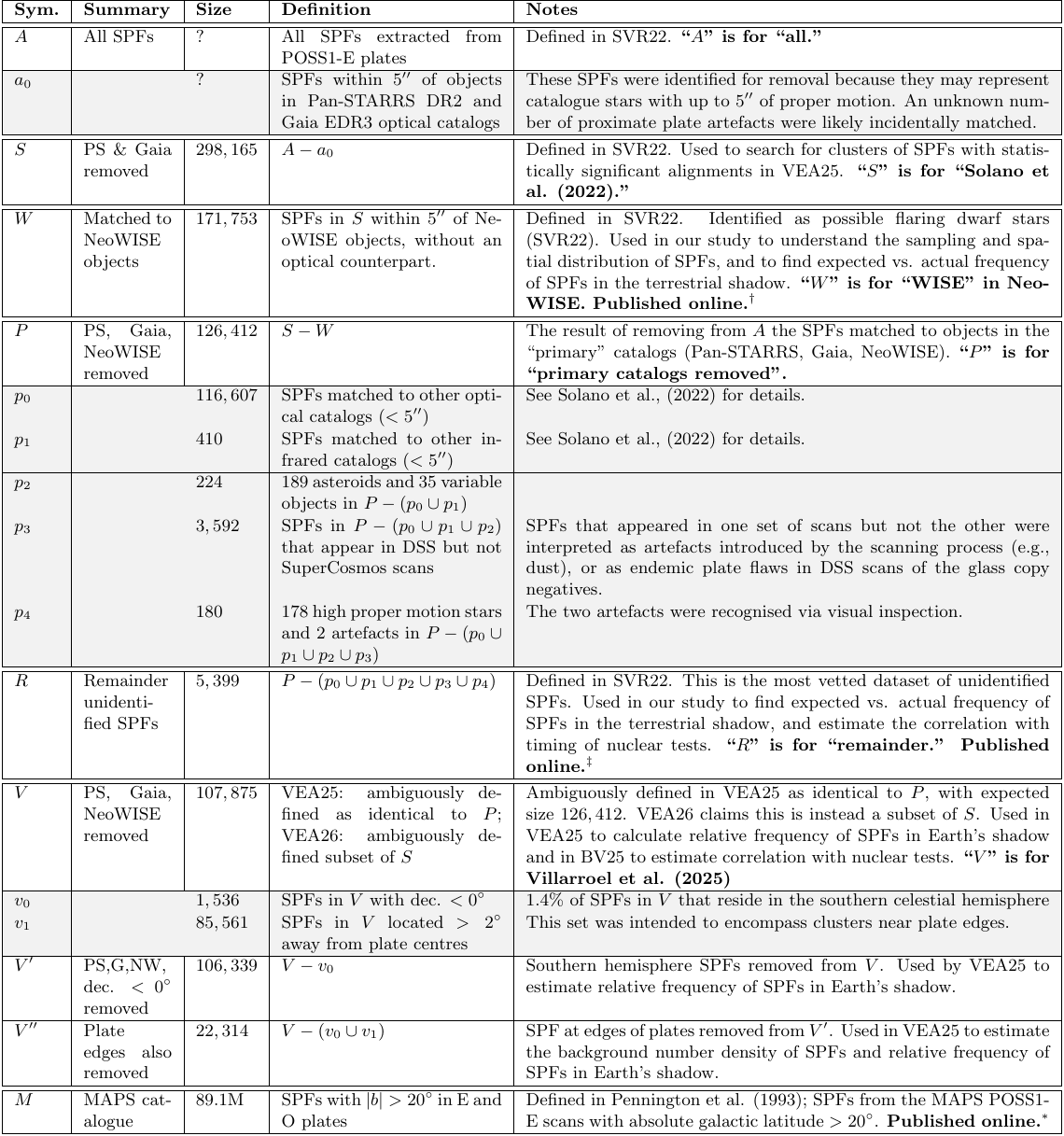}
\caption{Datasets analysed or discussed in the text.  These include datasets of Selected POSS1-E Features (SPFs) defined in \cite{solano2022discovering} (SVR22) and \cite{villarroel2025aligned} (VEA25) and also used in \cite{bruehl2025transients} (BV25).  A clarification about set $V$ appeared in \cite{villarroel2026response} v2 (VEA26).  Also defined is the $M$ dataset of likely celestial objects measured from the POSS1-E and POSS1-O plates described in \cite{pennington1993automated} and \cite{cabanela2003automated}.  Note that we have used these abbreviations: PS $\equiv$ Pan-STARRS, G $\equiv$ Gaia, and NW $\equiv$ NeoWISE. We have greyed the rows defining datasets with lowercase letters, which are used in the construction of the primary datasets (capital letters).}\label{tab_data_sets}

%\begin{tabnote}
{$\dagger$}{Retrieved on 2025-12-08 from {\tt http://svocats.cab.inta-csic.es/vanish-neowise/}}\\
{$\ddagger$}{Retrieved on 2025-12-08 from {\tt http://svocats.cab.inta-csic.es/vanish-possi/}}\\
{*}Retrieved on 2025-12-08 from {\tt https://aps.umn.edu/catalog/download/}\\
%\end{tabnote}
\end{table*}

\subsection{Summary of principal datasets}
The principal datasets discussed in our study are as follows:

\begin{itemize}

\item
{\it POSS1-E plate scans}: Publicly available plate images and metadata related to the POSS1 plates, which are the basis for all datasets described here. \cite{solano2022discovering} used a software pipeline to measure the positions of SPFs from these scans.  These scans and metadata are used in our study to investigate specific claims made in \cite{solano2022discovering} and \cite{villarroel2025aligned} about artefact removal, and in \cite{bruehl2025transients} about correlations between SPF occurrence and nuclear tests.

\item
{\it Set A}: Unpublished dataset from \cite{solano2022discovering}, of unknown size, comprising the initial extraction of SPFs from the POSS1-E plate set.

\item
{\it Set S}: Unpublished dataset from \cite{solano2022discovering} derived from set $A$ after removing optically visible catalogue stars (Gaia, Pan-STARRS), and used in \cite{villarroel2025aligned} to search for clusters of aligned SPFs. 

\item
{\it Set W}: Published dataset from \cite{solano2022discovering} containing SPFs from set $S$ that are near NeoWISE infrared catalogue object locations. This was used in our study to examine alignments, spatial distributions of SPFs, and for reevaluation of correlations with nuclear tests and Earth's shadow.

\item
{\it Set R}: Published dataset from \cite{solano2022discovering} used by our study as the most vetted dataset described in \cite{solano2022discovering}, \cite{villarroel2025aligned}, and \cite{bruehl2025transients}. This is used in the present work to identify artefacts, examine spatial distributions of SPFs, and to revaluate the Earth shadow and nuclear test correlations.

\item
{\it Set V}: Unpublished dataset derived from set $S$ and used by \cite{villarroel2025aligned} to explore SPF count deficits in Earth's shadow, and used by \cite{bruehl2025transients} to explore correlations between SPFs and nuclear tests.

\item
{\it Set M}: Published celestial catalogue from POSS1 plates by an independent research team, used in this study to examine characteristics of the spatial distribution of SPFs in $R$ and $W$ that are not apparent in $M$: i.e., that cannot be attributed to variations in the sensitivity of the optical system, the distribution of known celestial objects, or the astronomical viewing conditions on any given POSS1 observation day.  

\end{itemize}
 
\subsection{Dataset definitions}\label{sec_dataset_defs}
{\it POSS1-E}: The POSS1-E plates comprise 937 individual exposures using the red-sensitive Kodak 103a-E plate emulsion, whose sensitivity peaks at 640 nm with a limiting magnitude of 20 \citep{minkowski1963national}. This optical survey ran from  November 11, 1949 to December 10, 1958 and covered the whole sky north of declination -34$^{\circ}$. The 35.5 cm x 35.5 cm glass plates each have a $6.6^{\circ} \times 6.6^{\circ}$ field of view, and during the survey averaged 0.3$^{\circ}$ of overlap on each edge; \cite{pennington1993automated} and \cite{villarroel2025aligned} recorded the positions of features in a footprint with dimensions of roughly $6^{\circ} \times 6^{\circ}$, which is the approximate
area usable for measurements following the scanning process. This survey captured exposures for 292 plates with footprints residing fully in the southern celestial hemisphere; 584 plates with footprints residing fully in the northern celestial hemisphere; and 61 with footprints that overlap the celestial equator. The POSS1-E metadata contains the observation details such as start date/time, exposure, plate centre, and hour angle for each plate that is required for all temporal correlation analyses.  The DSS and SuperCOSMOS POSS1-E scans were made from two independent sets of glass copy negative plates made from a single set of glass copy positive plates, in turn made from the original glass negatives \citep{hambly2024nature}.

{\it Set $A$ (unknown size)}: \cite{solano2022discovering} began with the DSS scans of the POSS1 survey and extracted an unpublished dataset $A$ (see Figure \ref{fig_dbvenn}) whose size was not reported, containing all SPFs that passed their initial automated detection pipeline including signal-to-noise thresholds, spike removal, high proper motion, and morphological symmetry constraints (see \cite{solano2022discovering}).

{\it Set $S$ (N=298,165)}: To efficiently remove known stars while accounting for proper motion between surveys, each SPF in $A$ within 5$^{\prime\prime}$ of a catalogue object from Gaia DR3 \citep{vallenari2023gaia} or Pan-STARRS DR2 \citep{chambers2019panstarrs1surveys} was removed from $A$ to obtain the subset $S$ \citep{solano2022discovering}. For the set of SPFs removed from $A$, it was not reported whether there was a 1:1 correspondence between the catalogue object locations and SPFs. Those plates whose footprints reside entirely within the southern celestial hemisphere were never sampled, or else SPFs from those plates were removed at this stage.

{\it Set $W$ (N=171,753)}: Dataset $W$ is the publicly available subset of $S$ containing features located within 5$^{\prime\prime}$ of NeoWISE infrared catalogue objects \citep{mainzer2011preliminary} that are not visible in the optical catalogs, and which are not expected to be visible in POSS1-E except in cases where these objects are flaring red dwarfs \citep{solano2022discovering}.  We find a 1:1 correspondence between NeoWISE object locations and SPFs in $W$.  Therefore, $W$ is the result of using a spatially uniform-random distribution of positions (a starfield) to sample the set $S$, and is the focus of our analyses in Section \ref{sec_spatial_dist}. We determined that no SPFs in $W$ derive from plates whose footprints reside entirely within the southern celestial hemisphere.  \cite{solano2022discovering} then filtered the set $S$ by removing the set $W$, creating the intermediate set $P (N=126,412)$ (i.e., $P\equiv S-W$). 
 
{\it Set $R$ (N=5,399)}: In a series of stages, \cite{solano2022discovering} removed SPFs from $P$ to produce the published dataset we that have called set $R$ for ``remainder''.  First, \cite{solano2022discovering} removed all SPFs residing within 5$^{\prime\prime}$ of stars in additional optical and additional infrared astronomical catalogs (as described in \cite{solano2022discovering}).  Then, likely asteroids and variable objects were removed.  The remaining $9,171$ SPFs were searched for digitization-related artefacts (such as imaged dust or loose emulsion particles) by comparing separate scans of separate glass copy negatives (DSS and SuperCosmos) of the same field; an additional $3,592$ SPFs ($\sim 40$\%) were identified and removed in this way.  We expect that this step also removed endemic faults and ageing-related SPFs (such as plate corrosion) found on the DSS copy negative but not found on the SuperCosmos copy negative.  On the other hand, this would not have removed plate defects whose source was the original glass plate or the copy positive glass plate.  Next, SPFs identified as high proper motion stars and as obvious artefacts (by manual inspection) were also removed.  The resulting set $R$ ($N=5,399$) contains the unidentified SPFs least likely to be celestial objects, non-star-like plate artefacts, or digitization artefacts, and comprises less than 2\% of the SPFs in $S$. 

{\it Set $V$ (N=107,875)}:  \cite{villarroel2025aligned} defined set $V$ as set $S$ with the NeoWISE proximate SPFs removed ($S-W$) and with SPFs in the southern celestial hemisphere removed; this appears to be the same definition as our $P \equiv S-W$ because $W$ contains no SPFs from plates whose footprints completely reside within the southern celestial hemisphere.  This cannot be correct, however, since the count of SPFs in $V$ is short by 18,537 in light of the sizes reported for $S$ and $W$ in \cite{solano2022discovering}. \cite{villarroel2026response} v2 has since reported that the definition of $V$ in \cite{villarroel2025aligned} was incorrect: instead, $V$ derives from set $S$ (which includes all of set $W$) with an undisclosed number of additional southern hemisphere SPFs, catalogue stars, and duplicate SPFs removed. Although the details of the sampling process have not been disclosed, a uniform sampling of $S$ to get $V$ would yield roughly 60,000 SPFs from $W$ in $V$, of 107,295 in total.

Next, it can be inferred that removing the SPFs in the southern celestial hemisphere (on the equatorial plates) from set $V$ yields set $V^{\prime}$ (N=106,339). Set $V^{\prime\prime}$ ($N=22,314$) is defined as $V^{\prime}$ with SPFs near the plate edges removed. Detailed motivations for these inferences are supplied in Sections \ref{sec_caveats} and Appendix \ref{app_data_sources}.

\cite{villarroel2025aligned} and \cite{bruehl2025transients} did not use $R$ in the analyses discussed therein. \cite{villarroel2025aligned} used four datasets in its analyses ($S$, $V$, $V^{\prime}$, and $V^{\prime\prime}$), which are generally not concordant with definitions in \cite{solano2022discovering} and are ambiguously defined.  Set $S$ is tied to its definition in \cite{solano2022discovering} by the reported size ($N=298,165$). Although \cite{bruehl2025transients} cited \cite{solano2022discovering} as the source of its primary dataset, $R$ was not used but instead the set $V$ from \cite{villarroel2025aligned} was used; this can be inferred from the reported size ($N=107,875$). 

{\it Set $M$}: Nearly 90 million SPFs were recorded in the POSS1-E and POSS1-O plates by the MAPS project. MAPS discarded all features that did not appear in both the E (red band) and O (blue band) plates, in order to avoid recording plate artefacts; the Galactic plane was also occluded ($|b|> 20^{\circ}$) to avoid features that blend together in crowded fields. We chose the MAPS project digitization for comparison because it was independent of the DSS and SuperCOSMOS digitizations and because it records only very confident detections of celestial objects.

Our study makes use of the following datasets, all of which are publicly available: (i) set $W$ ($N=171,753$), consisting of SPFs near NeoWISE object locations (and the associated sky coverage map)\footnote{\url{http://svocats.cab.inta-csic.es/vanish-neowise/}}; (ii) set $R$, the most highly vetted SPF subset from \cite{solano2022discovering} (and the associated sky coverage map)\footnote{\url{http://svocats.cab.inta-csic.es/vanish-possi/}}; plot of SPFs in set $V$ on the sky (Figure 1 from \cite{villarroel2026response} v2); set $M$, the celestial catalogue generated from POSS1 plates \citep{pennington1993automated,cabanela2003automated}\footnote{\url{https://aps.umn.edu/catalog/download/}}.  We have also used the POSS1-E plate metadata \citep{stsci0000plate}\footnote{\url{https://gsss.stsci.edu/skysurveys/SurveyPlateList.txt}}, as well as POSS1-E images\footnote{\url{https://archive.stsci.edu/cgi-bin/dss_form}} in DSS\footnote{\url{https://archive.stsci.edu/dss/acknowledging.html}} and SuperCosmos scans\footnote{\url{http://www-wfau.roe.ac.uk/sss/}} (shown in Figures \ref{fig_example_flaws} and \ref{fig_newspots-svr22}) and Pan-STARRS images\footnote{\url{https://catalogs.mast.stsci.edu/panstarrs/}} (shown in Figure \ref{fig_example_flaws}). Finally, we use the three sources cited in Bruehl and Villarroel (2025) for nuclear test dates \citep{USNukes, SUNukes, GBNukes}.

Figure \ref{fig_dbvenn} illustrates the relationship between the $A$, $S$, $W$, $P$, $V$, and $R$ datasets defined in Table \ref{tab_data_sets}, which are defined or implied in \cite{solano2022discovering}, \cite{villarroel2025aligned}, \cite{bruehl2025transients}, and \cite{villarroel2026response} v2. This also illustrates the ambiguity (dashed line and ``?'') concerning the amount of overlap between the publicly available set $R$, which is the most vetted dataset from \cite{solano2022discovering}, and $V$, which has not been publicly released as of the time of this writing. None of the above papers describe the relationship between $R$ and $V$. We stress that access to $V$ is not necessary for our evaluation of the analysis and claims in \cite{villarroel2025aligned} and \cite{bruehl2025transients}.

\subsection{Problems with data sources}\label{subsec_data_quality} \label{sec_caveats}

In this section, we describe issues which, in principle, affect all of the SPF datasets, as well as issues that are dataset specific.

\subsubsection{Cross-cutting issue: emulsion faults}

Based on the historical work on GRB optical counterparts described in Section \ref{sec_grbot_litreview}, we estimate an upper bound for the number of endemic plate emulsion defects in \cite{solano2022discovering}'s initial set $A$ by using historical experimental estimates of their rate of production. These experiments were conducted using plate emulsions distinct from those used in the POSS1 survey and the defects were manually identified (versus automatically identified using an image processing pipeline). These earlier studies showed that certain ubiquitous plate emulsion defects appear star-like \citep{varady1992background, greiner1987search}. For example, \cite{greiner1987search} examined six laboratory plates that were first exposed to diffuse light (without a starfield) to simulate the sky background illumination, and which were then developed.
Through visual inspection, they found endemic faults that could be mistaken for stars at an average rate of 0.066 per cm$^2$.  Furthermore, the number of emulsion defects per unit area was found to increase towards the edges of the plate.  Using a similar procedure on plates with a different emulsion, \cite{varady1992background} found a rate of 0.377 per cm$^2$. These rates would translate to 83 and 475 star-like faults per POSS1 plate (35.5 cm $\times$ 35.5 cm), respectively.  Since two sets of plate emulsions preceded the copy negatives that were compared to remove defects, we double these rates to estimate of the number of these artefacts that could reside in both the DSS and SuperCosmos scans. Since set $A$ is sampled from approximately 645 relevant POSS1-E plates, we estimate that $A$ could contain a population of roughly 107,000 to 612,000 star-like emulsion faults. The initial morphological filter in \cite{solano2022discovering} would not have removed these symmetrical features, and some unknown fraction would have been removed incidentally because of proximity to Gaia, PS, or NeoWISE catalogue objects. If the Kodak 103a-E plate emulsion behaves in a similar way to the emulsions evaluated in these experiments, then it is likely that endemic plate emulsion defects make up a significant fraction of the SPFs in $V$ ($N=107,875$).  

\subsubsection{Dataset-specific issues}\label{subsec_data_specific}
{\it POSS1-E plate scans}:
There is no mention in \cite{solano2022discovering} about the removal of SPFs or plates imaged in the southern celestial hemisphere, yet there are no SPFs from plates fully in the southern hemisphere in the public datasets $R$ and $W$, nor shown in their published coverage maps (available at their respective data sites listed in Sec. \ref{sec_dataset_defs}). This leaves unexplained why the southern hemisphere plates were not sampled or, if they were sampled, why these SPFs were removed and at what stage in the construction of sets $S,P,W,R$ and $V$.

{\it Set S}: \cite{solano2022discovering} removed $288,770$ SPFs from the dataset $S$ in the process of creating $R$ because these resided within 5$^{\prime\prime}$ of infrared and optical sources in multiple catalogs \citep{solano2022discovering}. For unstated reasons, \cite{villarroel2025aligned} reverted to using the dataset $S$ to search for clusters of aligned objects, reporting these in Table 3 of that study. We investigate these clusters in Section \ref{sec_alignments}. 

{\it Set $W$}: This set comprises the SPFs in $S$ that are within 5$^{\prime\prime}$ of NeoWISE infrared catalogue object locations \citep{solano2022discovering}.  It is not known what fraction of these SPFs are dim stars that are flaring versus artefacts that have been incidentally matched to NeoWISE infrared catalogue object positions. \cite{solano2022discovering} cited multiple challenges with estimating the fraction of flaring red stars, such as strong dependence on angular distance from the Galactic plane and the incommensurable datasets and parameters that have been used to estimate flaring rates in specific studies. It is beyond the scope of this work to estimate the number of flaring red stars captured in the northern celestial hemisphere of the POSS1-E survey. $W$ overlaps with $V$ by an undisclosed amount; a uniform-random sampling of $V$ from $S$ would imply that $V$ is about 57\% comprised of SPFs in $W$.

\begin{figure*}[h]
\begin{center}
    \noindent\includegraphics[width=38pc]{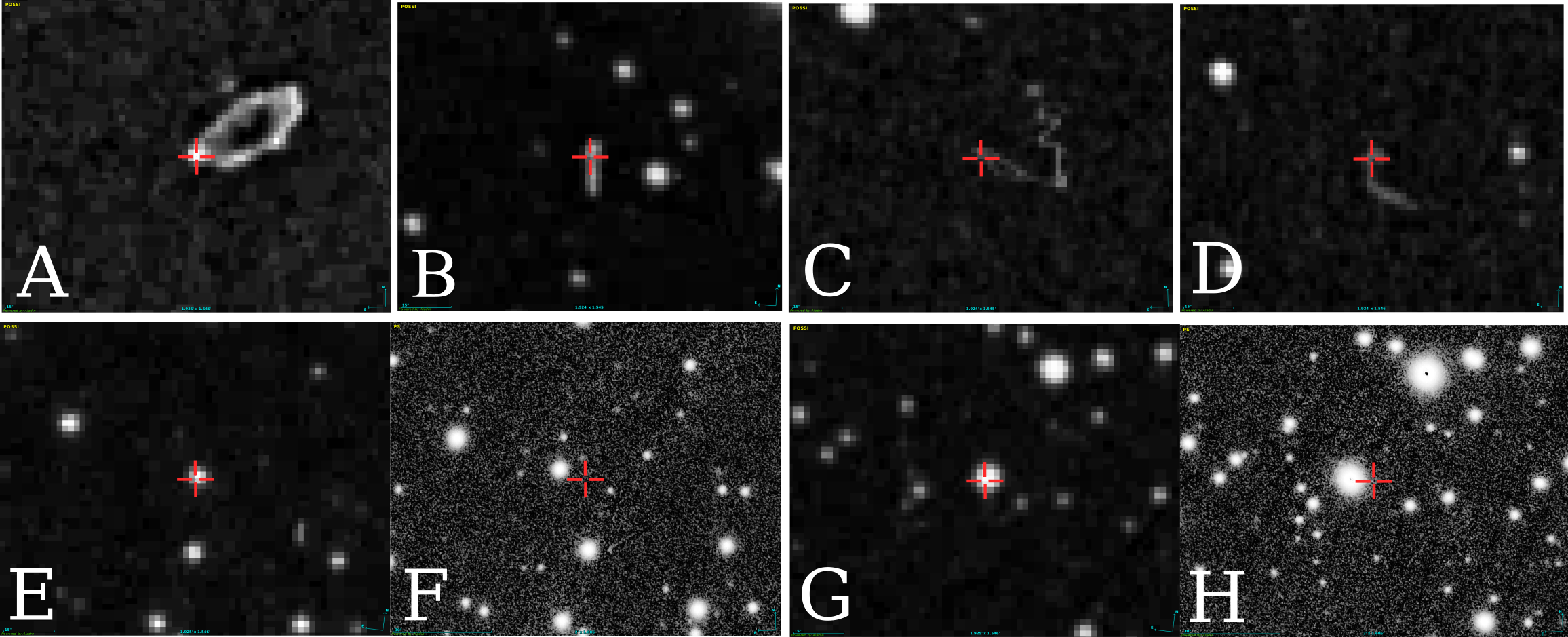}
    \caption{Examples of Selected POSS1-E Features (SPFs) in $R$ ($N=5,399$), the most vetted subset of $S$ ($N=298,165$) published by 
    \cite{solano2022discovering}.  The red crosses at the centre are $\sim 10^{\prime\prime}$ in diameter. We have identified these features as follows: (i) clear artefacts or defects that do not resemble objects visible in a Pan-STARRS image of the same field (A (object 1092), B (object 1179), C (object 1223), D (object 1248)); and (ii) clear examples of an object that was incorrectly identified as having vanished (object 1043 in E vs. F, and object 1051 in G vs. H, where E and G are POSS1-E images and F and H are Pan-STARRS images), probably on account of displacement from high proper motion. Based on 540 objects examined, cases of this sort comprise 4--5\% of $R$ ($N = 5,399$).  The images are available online from {\tt http://svocats.cab.inta-csic.es/vanish-possi/}.}
    \label{fig_example_flaws}
\end{center}
\end{figure*}

{\it Set $R$}: This set is the tiny residuum that survived the \cite{solano2022discovering} filter pipeline, which was designed to first remove all known celestial objects, and then digitization-related artefacts, from $S$. As the most highly filtered subset of $A$, we subjected $R$ to closer scrutiny by manually examining cropped POSS1-E images and Pan-STARRS images of the same field; 
in what follows, we supply ID numbers for the filenames of downloadable images in the public repository (see Section \ref{sec_dataset_defs}). We reviewed 10\% of the SPFs in $R$ (i.e., 540 images), finding 12 that appear to be of objects with exceptionally high proper motion (i.e., exceeding the $5^{\prime\prime}$ search radius) that appear in the Pan-STARRS frame (1043, 1051, 1059, 1130, 1243, 1271, 1277, 1345, 1374, 1376, 1409, 1513), and 11 that appear to form part of an obvious artefact (1092, 1248, 1255, 1460, 1179, 1223, 1259, 1301, 1396, 1473, 1507). Examples of these are shown in Figure \ref{fig_example_flaws}. We therefore estimate that between 4\% and 5\% (i.e., $\sim 23/540$) of the objects in $R$ are catalogue stars or obvious artefacts, which suggests that the overwhelming majority are roughly symmetrical SPFs without a catalogue match. 

{\it Set $V$}: Given that set $R$ was the most vetted of the \cite{solano2022discovering} datasets of unidentified SPFs, with as many known objects removed as then feasible, it was not clear why \cite{villarroel2025aligned} and \cite{bruehl2025transients} resampled $S$ to create a new set $V$ ($N=107,185$). \cite{villarroel2026response} v2 has since characterised the filtering in \cite{solano2022discovering} as overzealous in the context of searching for technosignatures, and was appropriate only for the case of searching for vanished stars.  It remains unclear why removing all possible confounding objects and artefacts is not also necessary in the search for technosignatures. 
Also, $V$ has not passed through the \cite{solano2022discovering} scan artefact removal step. We estimate up to $\sim40$\% of the objects in $V$ would be classified by \cite{solano2022discovering} as scan artefacts. This number is based on the size of the set preceding the scan filter ($N = 9,171$) and the size of the resulting set $R$ ($N = 3,592$), as reported in that study and detailed in Table \ref{tab_data_sets}.

\cite{villarroel2025aligned} used $V$ (or $V^{\prime}$) and $V^{\prime\prime}$ to search for correlations between SPF number density and the terrestrial shadow.  \cite{bruehl2025transients} also used $V$ to look for correlations between the days on which SPFs were recorded on POSS1-E plates and days with nuclear tests. More discussion concerning the ambiguous definition of $V$ can be found in Section \ref{sec_discussion} and Appendix \ref{app_data_sources}.

{\it Sets $W$, $R$, and $V$}: Public datasets $W$ and $R$ were sampled from plates whose footprints completely reside within or overlap the northern celestial hemisphere ($N = 645$). In addition, Figure 1 from \cite{villarroel2026response} v2 confirms that $V$ also contains no SPFs from plates fully in the southern celestial hemisphere. We infer that \cite{villarroel2025aligned} removed SPFs in the southern celestial hemisphere from equatorial plates ($N=1,536$ or 1.4\%) from $V$ ($N=107,875$) to arrive at $V^{\prime}$ $(N=106,339)$.  The fraction of SPFs residing below the celestial equator from equatorial plates in $R$ is also 1.4\%, and in $W$ is 0.07\%.  This is important because: (i) this supports our inference about the size of set $V$ in relation to set $V^{\prime}$ (with southern-hemisphere SPFs removed) reported by \cite{villarroel2025aligned}; (ii) it can be inferred that \cite{solano2022discovering} did not sample from plates whose footprints reside entirely within the southern celestial hemisphere when constructing set $A$, since there is no mention of removing southern hemisphere SPFs in that paper; and (iii) since neither $W$, $R$, nor $V$ include SPFs from plates fully in the southern celestial hemisphere, we are able to infer the correct number of relevant observation days within the study window when evaluating the reported correlation between SPF occurrence and nuclear tests (see Section \ref{sec_nuclear}), which differs from that used by \cite{bruehl2025transients}. We have used $W$ and $R$ to assist in reevaluating the purported correlations related to the Earth's shadow and nuclear tests (see Sections \ref{sec_shaddow_freq} and \ref{sec_nuclear}, respectively).

\section{Spatial distribution}\label{sec_spatial_dist}

In this section, we consider five aspects of the spatial distribution of the SPFs in $R$ and $W$ that provide information about their origin: (i) the variation in SPF counts per plate; (ii) their distribution with respect to the boundaries of photographic plates as a function of displacement from the plate centres; (iii) the nonrandom spatial distributions of SPFs on some plates, in which SPFs tend to avoid well-defined zones or reside mainly within simple geometrical shapes; and (iv) the SPF distribution across the sky.  Finally, we also address (v) the sporadic alignments of SPFs reported in Table 3 of \cite{villarroel2025aligned}.

Before proceeding, we define the phrase ``spatially uniform-random distribution'' or simply ``uniform-random'' as referring to the spatial distribution resulting from a Poisson point process: e.g., the spatial distribution characteristic of star fields.  We define ``nonrandom'' as any spatial distribution of features that, in principle, can be distinguished from ``uniform-random'' using a statistic like the Clark-Evans ratio \citep{clark1954distance}.   

\subsection{Variation in counts per plate}\label{sec_count_variation}

We examine the variation of counts per plate as part of assessing assumptions implicit in the analyses of \cite{villarroel2025aligned} and \cite{bruehl2025transients} about what factors supposedly influence the distribution of SPF counts in time and space, such as the terrestrial shadow and nuclear tests.  Primary sources of variation in counts per plate should be characterised, modelled, and subtracted {\it before} second-order effects are considered.   In particular, we might expect the number of SPFs per plate to vary for two primary reasons.  First, if the SPFs represent a population of stars not removed by prior filtering, then we expect (i) a roughly uniform-random distribution with an overall increase in number density with decreasing distance from the Galactic plane; and (ii) slight variations associated with variations in the astronomical viewing conditions (i.e., more stars are detected under favourable atmospheric conditions with less turbulence blur). Alternatively, if the SPFs represent plate artefacts, the distribution will depend on the processes that tend to create these features, and upon how the impact of these processes varies from plate to plate.  This may in turn depend on how the plates were manufactured, exposed, developed, stored, and handled, and how often they have been accessed. Finally, we should consider that the digital scanning process may also introduce artefacts, such as by introducing images of dust or debris in the scanner or on the plate surface.  \cite{solano2022discovering} removed digitization artefacts only after removing over 117,000 SPFs from set $P$ ($N$=126,412) when constructing set $R$ ($N=5,399$), whereas the removal of digitization artefacts was neither explicitly described nor mentioned in \cite{villarroel2025aligned} or \cite{bruehl2025transients} as part of the sampling of $V$ ($N=107,875$) from $S$.

Since the published tables for $R$ and $W$ do not list the plate of origin for each SPF, we have determined this as follows: for each SPF, we find all plate centres within 3$^{\circ}$ in declination, and then select the plate whose centre is nearest in horizontal separation.  We have noted all SPFs whose positions are overlapped by the fields of view of multiple plates, and which cannot be unambiguously assigned to a single plate, so that we can remove these in the analyses discussed in later sections.  This corresponds to about 10\% of the SPFs in $R$ and 12\% in $W$.

\begin{figure}[h]
\begin{center}
\noindent\includegraphics[width=20pc]{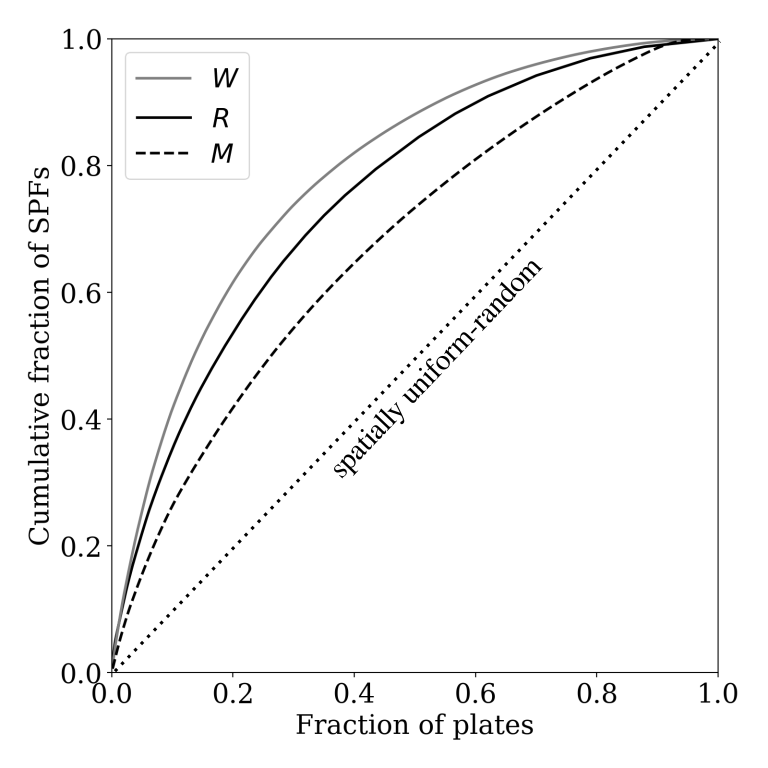}
\caption{Cumulative distribution function of SPF counts as a function of the total fraction of plates, ranked in order of decreasing counts from left to right, for $W$, $R$, and $M$.  In the case of $W$, roughly 20\% of the most crowded plates contain over 60\% of all SPFs.  $M$ is more uniform in terms of how SPFs are distributed across plates; this is partly because all SPFs in $M$ are likely celestial objects, and because the Galactic plane was occluded when this dataset was sampled. The curve still represents a significant departure from a uniform-random distribution (dotted line), possibly associated with variations in astronomical viewing conditions. The discussions of spatial variability in Sections \ref{sec_boundary_dist} and \ref{sec_templating} suggest there may be multiple factors contributing to the spatial variation in SPF counts shown in $R$ and $W$, arising from (i) how plates are manufactured, stored, scanned, and handled, or (ii) how datasets were constructed in \cite{solano2022discovering}.}
\label{fig_plate_counts_cdf}
\end{center}
\end{figure}

We found that counts of SPFs per plate in both $W$ and $R$ exhibit enormous variation.  In $W$, the number of SPFs per plate ranges from zero to 2,149 (plate ID 090R, label E1465), with an average of 257.5 and standard deviation of 337.2.  In $R$, the number of SPFs per plate ranges from zero to 105 (Plate 08BW), with an average of 8.1 and standard deviation of 10.3.  Figure \ref{fig_plate_counts_cdf} shows the cumulative fraction of SPF counts as a function of the total fraction of plates, ranked in order of decreasing counts from left to right.  The first datum plots as $\sim 0.012$ on the $y$ axis because the first plate, being the most crowded, comprises over 1\% of all the SPFs in the dataset.  This figure illustrates that 50\% of the plates contain nearly 90\% of the SPFs in $W$.  

We display the same plot for the $M$ dataset in Figure \ref{fig_plate_counts_cdf}, which shows a more gradual increase across the sample, but which still represents a significant departure from a uniform-random distribution (dotted line), possibly on account of variations in astronomical viewing conditions. Recall that all of the SPFs in $M$ correspond to likely celestial objects, as these are apparent in both the E and O plates (red and blue bands). Another potential reason for the difference in distributions is that whereas $M$ omits the bright zone within 20$^{\circ}$ of the Galactic plane, $S$ and its subsets (e.g., $R$ and $W$) were sampled from throughout most of this region.  That is, it is possible that the variation in SPF counts per plate in $S$ has been more strongly affected by the increased number density near the Galactic plane.  This could happen if (i) the SPF count comprises previously uncatalogued stars (e.g., flaring red dwarf stars), or (ii) the high number density of stars in this region result in blended objects with inaccurate centroid positions that frustrate the identification and removal of catalogue objects.

\subsection{Distribution with respect to plate boundaries}\label{sec_boundary_dist}

We will focus in this section on spatial distribution of SPFs within plates in set $W$ ($N = 171,753$) because (i) it is publicly accessible; (ii) it is large enough to clearly see patterns in relation to the plate centre and boundaries; and (iii) it is a majority subset of the set $S$ used by \cite{villarroel2025aligned} to search for linear clusters. Moreover, (iv) it will have retained some characteristics of $S$, the parent dataset of $V$, which is the focus of the analyses in \cite{villarroel2025aligned} and \cite{bruehl2025transients}.  This is because $W$ is effectively a dense, uniform-random sampling of the set $S$; i.e., it consists of SPFs residing within 5$^{\prime\prime}$ of NeoWISE infrared catalogue object positions, which are not expected to be visible in POSS1-E plates unless they are flaring red dwarfs \citep{solano2022discovering}. \cite{solano2022discovering} did not estimate the fraction of unidentified features consisting of actual flaring dwarf stars on account of the difficulty involved.  This is not necessary for our analysis, however, since it suffices for our purposes to show that the spatial distribution of SPFs in $W$ (and by implication $S$ from which $W$ and $V$ were derived) departs dramatically from the spatially uniform-random distribution expected of a starfield (such as a catalogue of NeoWISE objects), which has been assumed in the shadow deficit calculation in \cite{villarroel2025aligned}.  In summary, we focus here on the published dataset $W$ because of what we can learn about characteristics of the unpublished sets $S$ and $V$.

\begin{figure*}[h]
\begin{center}
\noindent\includegraphics[width=40pc]{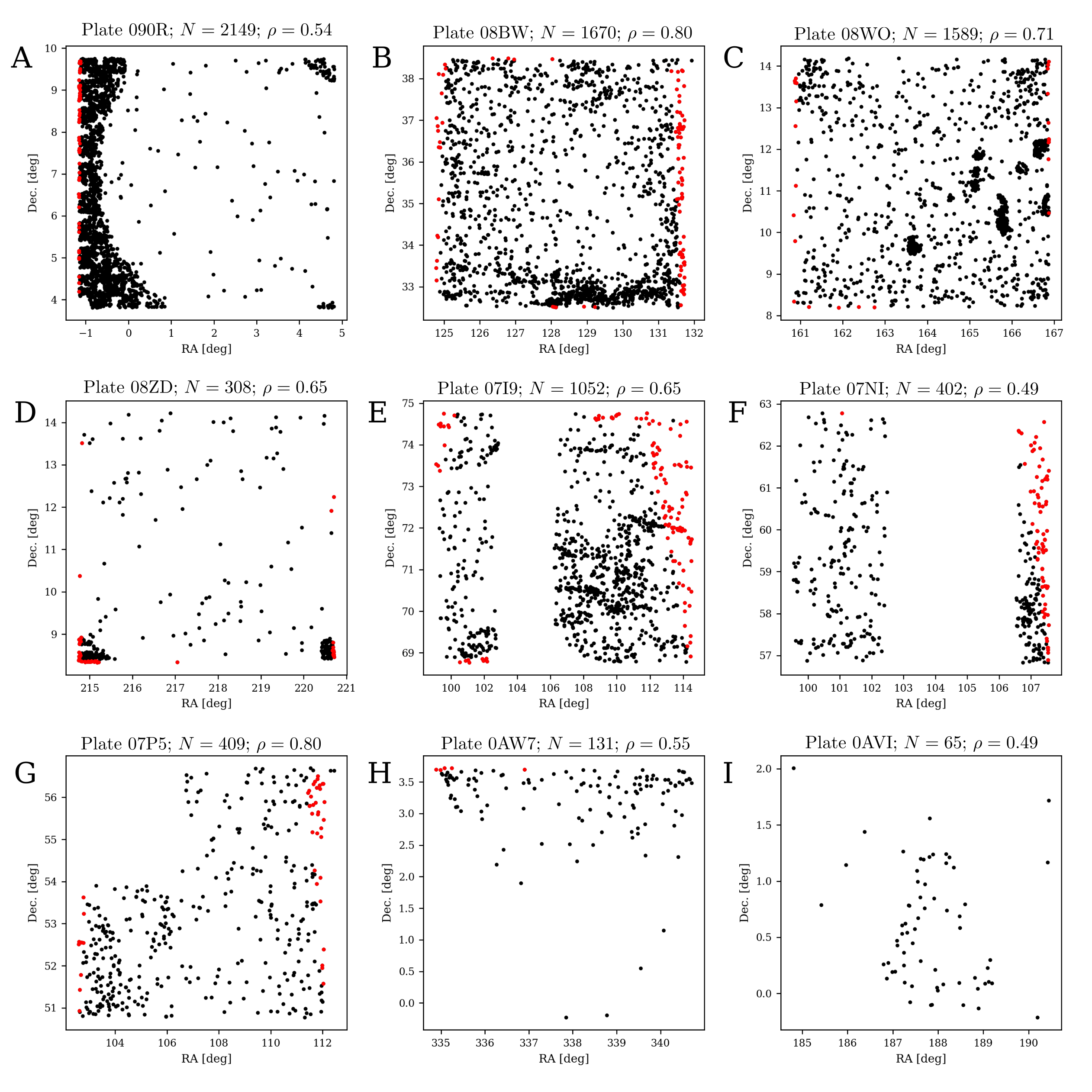}
\caption{Scatter plots of SPFs from $W$ on nine plates that exhibit distinctive patterns.  Red dots signify SPFs that could not be assigned to a single plate because of overlap with the neighbouring plate's field of view.  These patterns are not visible in the dataset $M$ of SPFs that appear in both E and O plates (i.e., likely celestial objects), suggesting that they are unrelated to nonuniform sensitivity of the instrumentation or photographic medium.  The geometric patterns and relationship to plate boundaries strongly suggest (i) a physical process that has templated the spatial distribution of SPFs, such as a surface that has come into contact with the plate, or a region that has been preferentially exposed to processes that tend to degrade the medium over time; or (ii) a spatial filter applied when processing the data.  We have plotted SPFs from $W$ because (i) it is publicly accessible; (ii) it is large enough for these patterns to be visible, and (iii) because it is a dense, uniform-random sample and majority subset of $S$, the parent set of set $V$ analysed by \cite{bruehl2025transients} and \cite{villarroel2025aligned}.  Each plate is labelled with the plate ID, the number of SPFs on the plate, and the value of the Evans-Clark ratio $\rho$, which is smallest for the most overdispersed (clustered) distributions. From left to right, the plate labels in each row are: E1465, E648, E976 (top); E65, E665, E1264 (middle); E971, E364, E1405 (bottom).}
\label{fig_unusual_patterns}
\end{center}
\end{figure*}

In Figure \ref{fig_unusual_patterns}, we have plotted the positions of SPFs in $W$ on selected plate fields that exhibit distinctive patterns, which in some cases bear a clear relationship to the plate boundaries.  We have marked the SPFs with ambiguous plate assignments in red.  These plates contain a wide range of SPF counts, most of them significantly exceeding the average of $W$'s 270 SPFs per plate.  All SPFs in $W$ reside within 5$^{\prime\prime}$ of NeoWISE infrared catalogue star positions and were removed from $S$ when constructing $R$ for this reason \citep{solano2022discovering}.
As Figure \ref{fig_unusual_patterns} shows, and as we demonstrate quantitatively later in this section, the distribution of SPFs from $W$ is not spatially uniform-random, which is the expected spatial distribution of a star field.
This suggests that the underlying distribution of SPFs in the parent set $S$ is also strongly nonrandom and also bears a clear relationship to the plate boundaries. 
Furthermore, the fact that the intra-plate distribution of SPFs in $W$ is far from uniform-random suggests that the large search radius (5$^{\prime\prime}$) and large number density of NeoWISE objects has resulted in sampling many noncelestial SPFs that were incidentally matched.  

Figure \ref{fig_unusual_patterns} reveals that there is a strong spatial variation within individual plates that is distinct from the plate-to-plate variation in astronomical viewing conditions and proximity to the Galactic plane.  Notably, on plates with large SPF counts, these appear to be sometimes concentrated in clusters.  To characterise the spatial distribution of SPFs within plates in $W$, we have calculated the Clark-Evans ratio ($\rho$) for each plate \citep{clark1954distance}.  This is defined as
\begin{equation}
\rho \equiv \frac{\bar{r}_{\rm obs}}{\bar{r}_{CSR}},
\end{equation}
where $\bar{r}_{\rm obs}$ is the observed average nearest neighbour distance between SPFs in a given plate, and $\bar{r}_{CSR}\equiv 1/{2\sqrt{\lambda}}$ is the average nearest-neighbour distance for a special case of the uniform-random distribution (generated by a Poisson point process) with average number density $\lambda$ \citep{clark1954distance}.  The special case is that of complete spatial randomness (CSR), defined as the spatial distribution resulting from a homogeneous Poisson point process on an infinite domain.  Because photographic plates have a finite extent, the nearest neighbour distance of SPFs near plate edges is larger than average, causing $\rho$ to be slightly more than 1 for points that are spatially uniform-random.  The case $\rho \gg 1$ implies a spatially underdispersed distribution (relatively regular spacing), while $\rho < 1$ implies overdispersion (clustering).

\begin{figure}[h]
\begin{center}
\noindent\includegraphics[width=20pc]{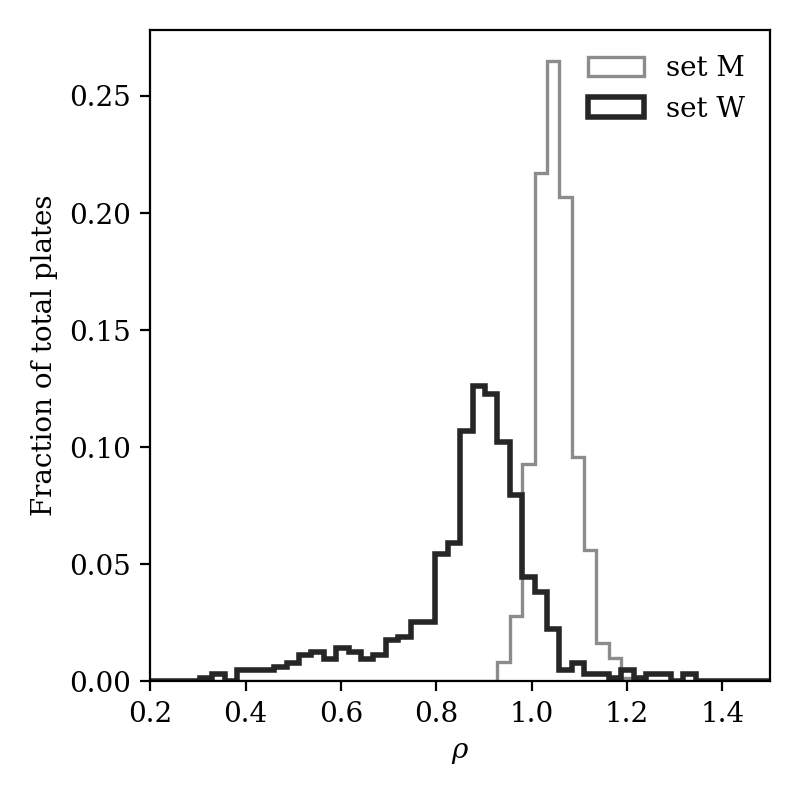}
\caption{Histograms of the Evans-Clark ratio $\rho$ calculated for each plate, for SPFs in sets $M$ and $W$.  As expected for a uniform-random distribution of points on a finite domain, $\rho$ is typically slightly above 1 for set $M$, comprising SPFs of likely celestial origin because they appear in both the POSS1-E (red band) and POSS1-O plates (blue band).  The distribution of SPFs in $W$ is dominantly overdispersed (clustered), with $\rho < 1$ for most plates.  See Figure \ref{fig_unusual_patterns} for examples of scatter plots of SPFs on individual plates with significant overdispersion, labelled with the estimated $\rho$.}
\label{fig_disp_hist1d_rho}
\end{center}
\end{figure}

We have computed $\rho$ on all of the plates containing SPFs for sets $W$ and $M$.  Set $M$ was downsampled to 0.2\% of its normal size to reduce computation time ($N \sim 170,000$), and so that its size is comparable to that of $W$ ($N = 171,753$).  Set $R$ has been omitted from this analysis on account of its small size, with the result that $\rho$ becomes a noisy statistic.  The scatter plots of SPFs in set $W$ for the plates in Figure \ref{fig_unusual_patterns} have been labelled with estimates of $\rho$, which ranges from 0.49 to 0.80 for these examples.  Figure \ref{fig_disp_hist1d_rho} contains a histogram of $\rho$ for plates in $M$ and $W$.  As expected, the spatial distribution of SPFs in the spatially uniform-random set $M$, consisting of SPFs that are very likely celestial in origin because they occur in both the O and E plates, usually result in values of $\rho$ slightly above 1 (modal value 1.04). By contrast, $\rho$ for plates in the set $W$ are widely distributed, peaking at modal value 0.89.  SPFs in $W$ are mostly overdispersed and exhibit moderate to significant clustering.  The scatter plots shown in Figure \ref{fig_unusual_patterns} are examples of significant overdispersion, sampled from the left-hand tail of the histogram in Figure \ref{fig_disp_hist1d_rho} for set $W$.

\begin{figure}[h]
\begin{center}
\noindent\includegraphics[width=20pc]{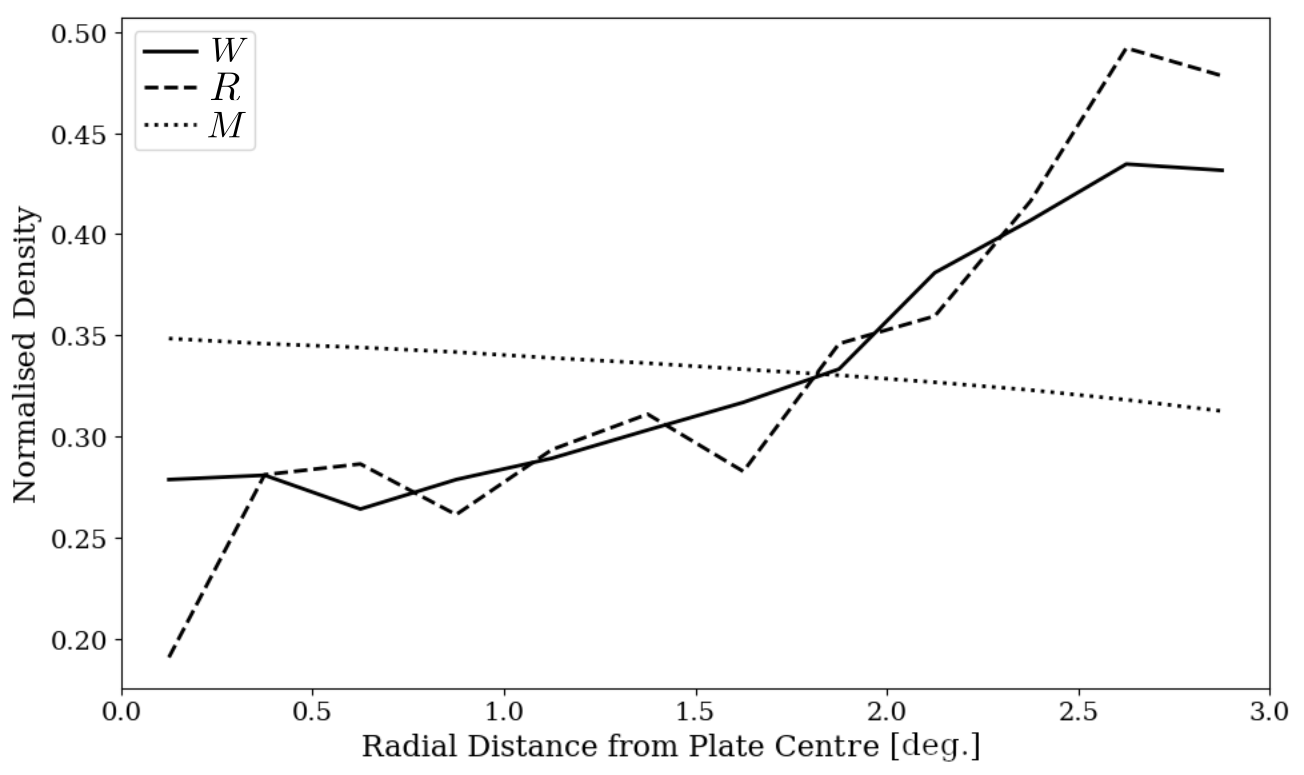}
\caption{Radial density profiles for the $W$ (solid), $R$ (dashed), and $M$ (dotted) datasets. Densities were calculated in circular bins of $0.25^{\circ}$ with radii denoting the angular distance from plate centres. All distributions were normalised for easy visual comparison of their shapes. The $W$ and $R$ distributions show an increasing density of SPFs away from plate centres, in qualitative agreement with the experimental results of \cite{greiner1987search}, who found that the number density of endemic plate emulsion defects increases with distance from plate centres. By contrast, $M$ exhibits a slight and gradual decline with increasing radius, indicating a mostly uniform number density combined with the effect of vignetting, as expected for a star field.  A $\cos({\rm Dec.})$ correction was applied to RA to correctly estimate east-west displacements.}
\label{fig_radial_density}
\end{center}
\end{figure}

\begin{figure}[h!]
\begin{center}
\noindent\includegraphics[width=19pc]{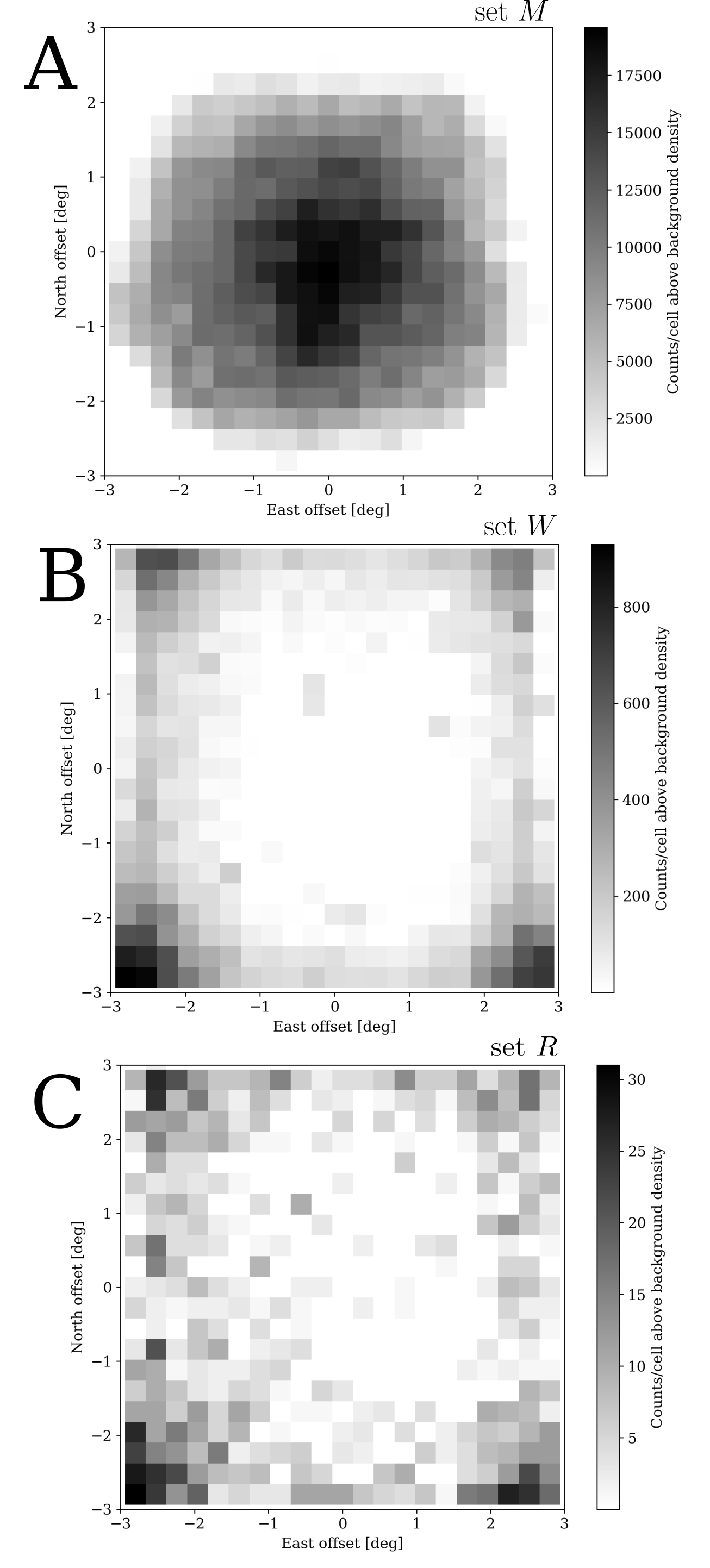}
\caption{Histograms (2-D) of SPF counts in excess of the median background as a function of offset from plate centres for (A) set $M$ (likely celestial objects that appear in both O and E plates; see \cite{cabanela2003automated}); (B) set $W$, which reside within 5$^{\prime\prime}$ of the positions of NeoWISE infrared catalogue objects \citep{solano2022discovering}; and (C) set $R$, comprising unidentified SPFs.  
The set $M$ distribution peaks in the centre, possibly on account of the effects of vignetting at the margins.  The distributions for set $W$ and $R$ peak in the edges and corners, and are qualitatively consistent with the expected pattern for endemic plate emulsion defects \citep{greiner1987search}. A $\cos({\rm Dec.})$ correction was applied to RA to correctly estimate east-west displacements.}
\label{fig_offsets_hist2d_x3}
\end{center}
\end{figure}

Next, we characterised the average spatial distribution within photographic plates.  For the $M$, $W$, and $R$ datasets, we computed the displacement in decimal degrees of SPFs with respect to the centre of each plate in declination and right ascension.  The right ascension was multiplied by the cosine of the declination to compute true east-west displacement on the sky.  These displacements were used to compute a normalised radial density profile in Figure \ref{fig_radial_density} for sets $M$, $W$, and $R$.  For each data set, we have also broken down the spatial distribution of the total SPF counts across all plate fields into a median background plus a nonrandom excess.  In Figure \ref{fig_offsets_hist2d_x3}, we have subtracted the median background SPF count within 2-D bins in order to visualise the excess SPF counts; this is shown in parts A, B, and C for sets $M$, $W$, and $R$, respectively.  In plotting parts B and C, we have used only the SPFs that can be unambiguously assigned to plate footprints.  

The distribution for set $M$ is consistent with expectations for telescopic observations of star fields: the centres of photographic plates tend to contain slightly more counts than the perimeter owing to the effect of vignetting near the margins.  By contrast, the excess in $W$ and $R$ shows the opposite pattern, with the number density of SPFs increasing with distance from the centre of the plates, and is therefore highest at the corners and edges.  
The amplitude of this radial increase, as a fraction of total counts, is also much larger than the amplitude of the variation in $M$ (Figure \ref{fig_radial_density}).  Recall that $R$ supposedly contains only features that are not identifiable as plate artefacts, digitization artefacts, or celestial objects, and so it is difficult to form an expectation in advance about the intra-plate spatial distribution.  By definition, all SPFs in $W$ are located in close proximity to the positions of NeoWISE objects. Nevertheless, $W$ exhibits a pattern that is inconsistent with the expectation of a uniform-random distribution for a star field on an individual plate: i.e., it does not resemble the distribution of SPFs in $M$.  

The observed pattern for $R$ and $W$ instead suggests a distinct generative process that depends on proximity to plate boundaries. As part of the GRB-related efforts to understand plate faults, \cite{greiner1987search} examined laboratory plates that had been exposed to diffuse light equivalent to sky background illumination, and which were then developed. Using different emulsions and plates, they found endemic faults at a rate that translates to 83 faults per area of a POSS1 plate (35.5 cm $\times$ 35.5 cm) that were difficult to distinguish from ideal stars or slightly distorted stars. These faults included point-like objects indistinguishable from stars, oblong objects indistinguishable from stars with trails, and objects with comas located on the outer edges of plates that resembled a normal star with slight distortion. The number of faults increased towards the edge of the plate and also with decreasing brightness. Other mechanisms for generating non-random excesses of SPFs on the edges of plates have not been experimentally examined; these include the contamination or degradation of the emulsion through handling of plates by the corners and edges. The distributions of SPFs shown here are plausibly related to the manufacture, processing, degradation, or digitization of the plates and unrelated to luminous or illuminated objects in the telescope's field of view.   

Finally, we highlight that in all of our analyses, a $\cos({\rm Dec.})$ correction was applied to RA to obtain accurate east-west displacements.

\subsection{Templating and spatial filtering}\label{sec_templating}

In some cases, the distribution of SPFs within plates exhibits patterns and clusters that are qualitatively distinct from the mean pattern described in the previous section.  A prominent example occurs very near the celestial meridian and equator in the plate with the largest number of SPFs (plate ID 090R, label E1465: $N$ = 2,149, in part A in Figure \ref{fig_unusual_patterns}), where the majority of SPFs are located between the boundaries of the plate and an oval that is partially inscribed within it.  The origin of these patterns is not known.  Other plates exhibit rectangular voids (part G) or  band-like vertical gaps (parts E and F).  We also find examples of tight clusters within central regions of a plate (part C) or at the corners (part D).  A few plates exhibit a density gradient with highest density at one edge, decreasing toward the opposite edge (part H).  

That some plates exhibit patterns bearing a clear relationship to plate boundaries (e.g., clustering near edges) and/or the celestial coordinate system (e.g., voids and clusters exhibiting horizontal and vertical edges, parallel to lines of constant right ascension or declination), suggests the action of generative mechanism(s) that relate to the manufacture, processing, curating, or digitizing of the plates, rather than to lights in the sky.  

The fact that some SPFs occur within regions with clear boundaries, or else avoid regions with clear boundaries, suggests that contact or exposure-based templating may play a role in the production of SPFs more generally.
An example zone of avoidance is visible in the upper-left corner in part G of Figure \ref{fig_unusual_patterns}.  Based on inspection of multiple plates, the vertical stripes in parts E and F span the interval in right ascension from roughly $103^{\circ}$ to $107^{\circ}$ and {\it cross} plate boundaries; a second stripe from roughly $86^{\circ}$ to $98^{\circ}$ also crosses plate boundaries.  Both of these features are further discussed in Section \ref{sec_sky_dist}.  

We emphasise the importance of answering the following questions. First, why do SPFs in some cases appear to reside in clusters or regions of higher number density with clear boundaries, and why do they sometimes appear to avoid regions with clear boundaries?  Second, why do these overdense regions and empty regions in some cases appear related to the plate footprint or the celestial coordinate system (e.g., plate ID 090R, label E1465 in part A of Figure \ref{fig_unusual_patterns})?  Third, what does this imply about the production of SPFs in these POSS1-derived datasets?  These questions should be answered as part of the essential data validation process.

\subsection{Distribution across the sky}\label{sec_sky_dist}

\begin{figure*}[h!]
\begin{center}
\noindent\includegraphics[width=37pc]{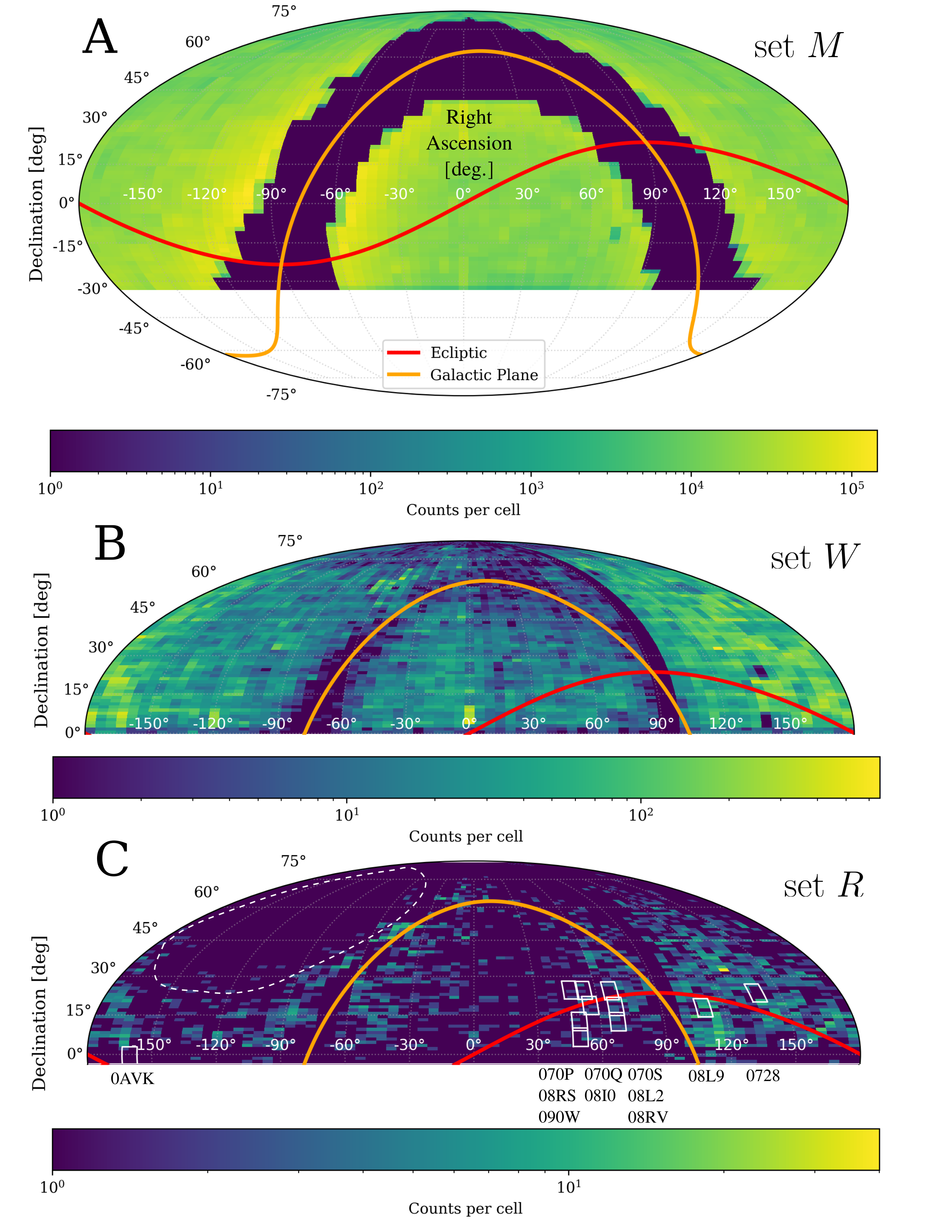}
\caption{Plots of the number density of SPFs on the celestial sphere in (A) set $M$, (B) set $W$, and (C) set $R$.  The Galactic plane was occluded in set $M$ ($|b|> 20^{\circ}$) to avoid features that blend together in crowded fields.  The distribution of SPFs in parts B and C reveal a vertical band with a deficit of SPFs between 90$^{\circ}$ and 105$^{\circ}$; this remains apparent following aggressive filtering in $R$. There is also a marked deficit in a band surrounding the plane of the Galaxy; this is not explained in \cite{solano2022discovering}. The white boxes in (C) show the outlines of plates estimated to overlap Earth's shadow during the observation window, as discussed in Section \ref{sec_shaddow_freq}; their plate IDs are supplied below the plot.  \cite{villarroel2025aligned} reported a deficit of SPFs from set $V^{\prime}$ within these plates. The red line shows the ecliptic, the orange line is the Galactic plane, and the dashed white line highlights a marked depletion north of the celestial equator in the western hemisphere when compared to the eastern hemisphere.}
\label{fig_sky_plots_x3}
\end{center}
\end{figure*}

In this section, we examine significant variation in SPF number density on the scale of the entire sky.  We have graphed the on-sky distribution of the number density of SPFs from set $M$ in part A of Figure \ref{fig_sky_plots_x3} (i.e., likely celestial objects that appear in both O and E plates); we have graphed the number density of SPFs from $W$ in part B and from $R$ in part C.  As mentioned, \cite{cabanela2003automated} occluded a buffer zone surrounding the Galactic plane to avoid blended objects in crowded fields when constructing $M$ ($|b| > 20^{\circ}$).  By contrast, \cite{solano2022discovering} does not specifically mention avoiding the Galactic plane while sampling.

We have shown the all-sky map of SPF number density in Figure \ref{fig_sky_plots_x3} to call attention to the significant variation in this quantity across the sky in the \cite{solano2022discovering} datasets, and especially the vertical stripe between roughly RA = 90$^\circ$ and RA = 105$^\circ$ for both $R$ and $W$.  Closer inspection of the plate-framed scatter plots (as shown in Figure \ref{fig_unusual_patterns}, parts E and F) suggests that this stripe consists of two bands, from roughly RA = 87$^{\circ}$ to RA = 98$^{\circ}$ and a second, narrower band from RA = 103$^{\circ}$ to RA = 107$^{\circ}$.  This stripe lies largely in the galactic occlusion zone of set $M$, but no such stripe is visible in scatter plots of SPFs in $M$ near the poles.  Therefore, this feature in $W$ and $R$ does not appear to have resulted from a failure of the instrument to faithfully record astronomical objects in this region.  There is also a notable band surrounding the Galactic plane, which is also not explained in \cite{solano2022discovering} in terms of a specific filtering or sampling process.
These large scale spatial heterogeneities are also clearly visible in the published coverage maps of $W$ and $R$ \citep{solano2022discovering}, as well as the scatter plot of SPFs in $V$ \citep[][v2]{villarroel2026response}. 

We considered four possible explanations regarding the origin of the stripes: (i) {\it The SPFs were incompletely sampled when the set $A$ was assembled by \cite{solano2022discovering} in the first place, or when the DSS or SuperCOSMOS scans were acquired.}  This appears unlikely because the stripes do not correspond to plate boundaries: they cross plate boundaries and only partially overlap some plates (see Figure \ref{fig_unusual_patterns}, parts E and F). Note that plate boundaries become staggered as the sky area narrows near the pole, and yet the stripes span the full range of declinations: they are not offset or shifted at plate boundaries.  Moreover, the striping did not affect set $M$, which was independently scanned; this suggests the stripes are not an intrinsic property of the plates themselves.  (ii) {\it The stripes are the result of how the plates were stored, maintained, and handled.}   This seems unlikely because it is hard to imagine how any of these activities could create a stripe that traces lines of roughly constant right ascension and which cross plate boundaries, especially given the staggered arrangement of plate footprints on the sky.  (iii) {\it The POSS1 telescope observations overlapping the stripe were captured during periods of time when SPFs, if they represent luminous or illuminated objects in the telescope's field of view, were less abundant in the sky.} This can be confidently ruled out because the plates overlapping the stripes were captured throughout the duration of the POSS1 survey (8 years).  Also, as mentioned already, the stripes only partially overlap the plates, ruling out a time dependence.  This leaves the least unlikely cause: (iv) {\it SPFs within the stripes were filtered out during the processing steps that led from set $A$ to sets $W$ and $R$ in \cite{solano2022discovering}}.  It is not clear why this interval would be singled out for suppression, intentionally or by accident.

We call attention to this feature of the on-sky distribution because analyses in \cite{villarroel2025aligned} and \cite{bruehl2025transients} implicitly assume that $V$ is completely and uniformly sampled (i.e., no part of the north celestial hemisphere was omitted when SPFs were selected) and faithfully records the distribution of SPFs across the sky.  This assumption is essential for one of the key claims in \cite{villarroel2025aligned}, which reports a deficit of SPFs from set $V^{\prime}$ in the Earth's shadow, as discussed in the next section.  This is also significant for interpreting the background rate of SPFs estimated in the same paper.  An incomplete or nonuniform sampling of the sky would affect correlation statistics between SPF occurrence and any other phenomena, such as nuclear tests in \cite{bruehl2025transients}.

There are marked similarities and differences evident from comparing the spatial distributions of $W$ and $R$ (see parts B and C of Figure \ref{fig_sky_plots_x3}), which shows the effect of removing additional catalogue objects and scan artefacts from $S$ by \cite{solano2022discovering}.  Across the western celestial hemisphere, SPFs in the set $R$ ($N = 5,399)$ tend to hug the plane of the Galaxy, suggesting that some fraction of these objects may be uncatalogued stars, including uncatalogued dwarf stars that were transiently flaring during the POSS1-E exposures.  Alternatively, the high number density of stars in this region result in blended objects with inaccurate centroid positions that may frustrate the identification and removal of catalogue objects.  There is also a marked depletion north of the celestial equator in the western hemisphere when compared to the eastern hemisphere in both $W$ and $R$ (white dashed outline in Figure \ref{fig_sky_plots_x3}).  In the case of $R$, SPFs in the western hemisphere above the Galactic plane have been almost totally erased.  
Before using any of these datasets, $R$, $W$, or $V$, to search for images of unidentified objects, the origin of these patterns in the spatial distribution should be investigated and understood.

\subsection{Clusters of aligned SPFs}\label{sec_alignments}

\cite{villarroel2025aligned} also searched for clusters of up to five SPFs exhibiting statistically significant alignments: i.e., that are unlikely to result from a Poisson point process for populations of comparable size.  These were called ``candidate alignments'' and were presented in Table 3 of that study.  As already mentioned, there is no reason to expect that plate artefacts that give rise to SPFs should have a spatially uniform-random distribution (as if generated by a Poisson point process). We presented strong evidence that they are not uniform-random distributed, such as by highlighting SPFs in regions with clear linear boundaries (Section \ref{sec_templating}). In light of these observations, discovering clusters of aligned SPFs is not especially surprising. In this section, we focus on the nature of the specific SPFs that have been identified as belonging to the candidate linear clusters in Table 3 of \cite{villarroel2025aligned}.

The start of Section 4 in \cite{villarroel2025aligned} states that: ``We base our analysis on the catalogue of 298,165 short duration transients presented in Solano et al. (2022), detected in red POSS1 plates with typical exposure times of 45–50 minutes.'' This implies that the search for aligned SPFs made use of $S$ ($N=298,165$), which according to \cite{solano2022discovering} is 96.8\% composed of features that reside within $5^{\prime\prime}$ of a catalogue star (including what we have called set $W$). Indeed, as shown in our Table \ref{tab_alignments}, we find that 8 of the 24 SPFs belonging to candidate linear clusters in Table 3 of \cite{villarroel2025aligned} reside within 2$^{\prime\prime}$ of an object in $W$ (a subset of $S$), suggesting that these features represent one and the same object. 
\cite{villarroel2025aligned} recomputed the astrometric positions of SPFs in Table 3 \citep{villarroel2025aligned} with respect to positions originally reported for SPFs in $W$ from \cite{solano2022discovering}, which likely accounts for the  $<2^{\prime\prime}$ positional differences we see here. We have confirmed that 8 of 24 SPFs in Table 3 of \cite{villarroel2025aligned} are located within 5$^{\prime\prime}$ of a NeoWISE catalogue object position (a 9th is $\sim 5.1^{\prime\prime}$ away). These members of set $W$ were removed from $S$ by \cite{solano2022discovering} when constructing $R$ because they were not confidently distinguished from NeoWISE objects. For the same reason, \cite{villarroel2025aligned} also claims to have removed $W$ from $S$ to construct the set $V$ (a claim since retracted in \cite{villarroel2026response} v2) used in other analyses. No reason is given in \cite{villarroel2025aligned} for reverting to set $S$ to search for SPF alignments representing satellite glints.

\begin{table}[]
\centering
\caption{SPFs within quasilinear clusters reported in Table 3 of \cite{villarroel2025aligned} with an angular separation ($d_{\rm W}$) of less than 2$^{\prime\prime}$ from an SPF in $W$.  That is, 8 of 24 listed are probable matches to SPFs in $W$.  Set $W$ consists of SPFs whose celestial coordinates at the time of observation resided within 5$^{\prime\prime}$ of a NeoWISE catalogue object.  RA and Dec. coordinates (sexag.) are reported for the J2000 reference frame.}\label{tab_alignments}
\begin{tabular}{lllll}
\hline\hline
Cluster & Object & RA & Dec. & $d_{\rm W}$ [as]\\
\hline\hline
1 & 2 & 02:29:21.38 & +28:36:57.89 & 1.7 \\
  & 3 & 02:29:21.76 & +28:36:49.09 & 1.8 \\
\hline
4 & 3 & 21:24:47.62 & +68:31:58.92 & 0.7 \\
  & 4 & 21:24:39.72 & +68:31:31.22 & 1.1 \\
  & 6 & 21:24:03.94 & +68:29:14.36 & 0.8 \\
\hline
5 & 3 & 19:16:45.73 & +51:28:52.04 & 1.8 \\
  & 4 & 19:16:40.13 & +51:27:12.85 & 1.7 \\
  & 5 & 19:16:40.27 & +51:27:06.29 & 1.7 \\
\hline\hline
\end{tabular}
\end{table}

\begin{figure*}[h]
\begin{center}
\noindent\includegraphics[width=37pc]{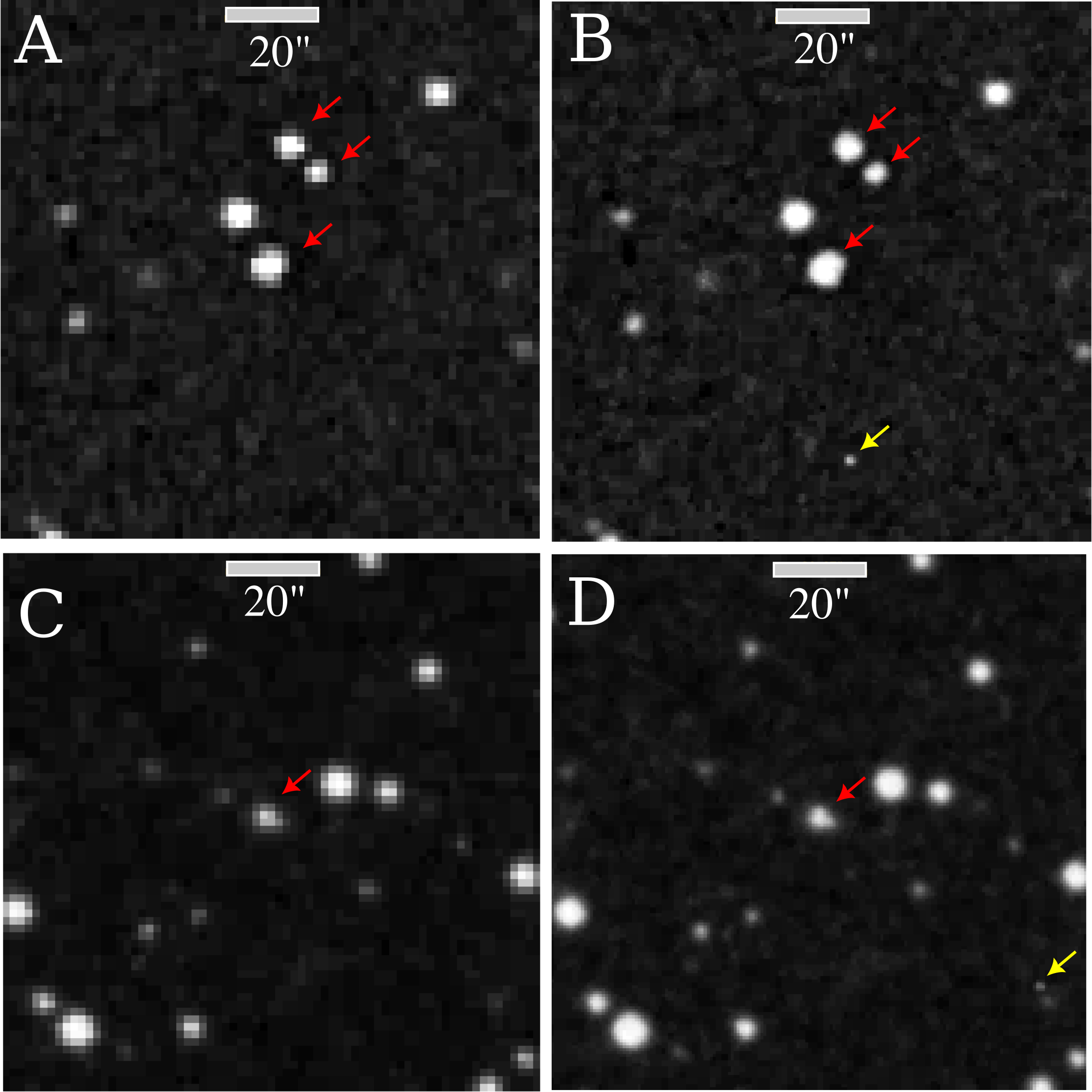}
\caption{Magnified images of SPFs belonging to candidate linear clusters detailed in Table 3 of \cite{villarroel2025aligned} (red arrows) and SPF scan artefacts present in SuperCosmos scans (B and D) and absent from DSS scans (A and C) of the same field (yellow arrows).  Parts A and B represent ``candidate 3'' and Parts C and D represent ``candidate 5''.  The noticeable departure from circular symmetry in the case of multiple objects marked with red arrows suggests that $S$ should be aggressively filtered to remove features without an extremely high degree of radial symmetry, as expected for the point source glints hypothesised in \cite{villarroel2025aligned}, in order to avoid likely artefacts.}  
\label{fig_newspots-svr22}
\end{center}
\end{figure*}

We manually examined the SuperCosmos and DSS images of all SPFs in Table 3 of \cite{villarroel2025aligned} and show two comparisons in Figure \ref{fig_newspots-svr22}.  We have noted a marked qualitative asymmetry of some of these SPFs, indicating the need for a quantitative morphological analysis in addition to the symmetry-based filtering already applied in \cite{solano2022discovering}.   (A morphological analysis is also recommended by the presence of the anomalous shapes of some SPFs in $R$ as shown in Figure \ref{fig_example_flaws}.) In Figure 8 of \cite{villarroel2025image}, the lowermost member of the cluster of three SPFs (red arrows in part B of our Figure \ref{fig_newspots-svr22}) is described as ``slightly dubious in shape''.  In our high-magnification view, this is clearly an asymmetrical feature and not representative of a  star-like optical transient. 
These images also highlight two defects of the kind that can be removed by scan comparison, in this case appearing on the SuperCosmos scan and not the DSS scan (yellow arrows).  These artefacts are also clearly identified as such in Figure 8 of \cite{villarroel2025aligned}. Parts C and D of our Figure \ref{fig_newspots-svr22} show another SPF from an aligned cluster (object 3 in cluster 5 of Table 3 in \cite{villarroel2025aligned}) that also exhibits clear elongation, although is possibly a pair of overlapping sources. 

\section{Frequency of occurrence in Earth's shadow}\label{sec_shaddow_freq}

\cite{villarroel2025aligned} have alleged there is a deficit of SPFs within the Earth's umbra, whose radius is roughly 8.5$^{\circ}$ at the altitude of geosynchronous orbits. This was reported as a $30-75\%$ reduction in the frequency of SPFs with respect to expectations, at levels of statistical significance ranging from $\sigma = 2.5$ to $\sigma = 22$.  The significance was calculated using Poisson uncertainties, which are appropriate for Poisson point processes: i.e., it was assumed that the background population of SPFs is spatially uniform-random, which we showed is invalid for set $W$ in Section \ref{sec_spatial_dist}.  The reported deficit of SPFs in the shadow was cited as evidence for the hypothesis that a significant fraction of SPFs represent objects reflecting sunlight in geosynchronous orbits. According to \cite{villarroel2025aligned}, such objects are expected to produce point-like glints (if sufficiently faceted and rotating fast enough) rather than along-track streaks during the POSS1-E $\sim$50 minute exposures.  As mentioned earlier, the study acknowledged the existence of invalid detections in the dataset $V$ (e.g., plate artefacts), but has assumed that these would exhibit a uniform-random distribution that is superposed on the true distribution of non-astronomical light sources.  In addition to the null statistical hypothesis just mentioned, this assumption is explicitly stated (e.g., see section 3 on ``Predictions and Expectations'' in \cite{villarroel2025aligned}: ``Plate defects, by contrast, are expected to be randomly shaped and distributed'').  

Elsewhere, \cite{villarroel2025aligned} acknowledges the overdensity of SPFs near plate edges without plotting, quantifying, or otherwise characterizing this effect.  
SPFs more than $2^{\circ}$ away from plate centres were removed by \cite{villarroel2025aligned} (i.e., to construct set $V^{\prime\prime}$; see Table \ref{tab_data_sets}) and the in-shadow SPF frequency analysis was repeated, still finding a significant deficit of SPFs in the shadow ($\sim 30$\%), although with marginal statistical significance owing to the smaller sample size ($\sigma = 2.6$). \cite{villarroel2025aligned} acknowledged the increase in SPFs near plate edges as plate-related, without demonstrating that the processes that generate SPFs near the edges of plates do not also generate SPFs at the centres of plates.  

\cite{villarroel2025aligned} relies on the results of an inferential analysis (a supposed deficit of SPFs in the terrestrial shadow) to draw a conclusions about the likely origin of the SPFs (glinting spacecraft) as well as to justify confidence in the validity of the measurements.  In other words, the results are used both as evidence that SPFs are optical flashes recorded in the emulsion, and as evidence that SPFs are glinting objects in the sky.  In Section \ref{sec_circular}, we argue that this is an example of circular reasoning. 

\subsection{Shadow Calculation}
To assess the reported deficit of SPFs in the Earth’s
shadow, we have simulated the shadowing of plates. Here we attempt to replicate the procedure by \citet{villarroel2025aligned} as closely as possible. Because set $V$ was not publicly available at the time of this writing, we used the $W$ and $R$ datasets, which were sampled from the same parent dataset as $V$ and which, based on their sky coverage maps from \cite{solano2022discovering} and in Figure \ref{fig_sky_plots_x3}, bear a strong resemblance to $V$ at large spatial scales (see Figure 1 in \cite{villarroel2026response}).

We model the Earth's shadow using the \texttt{earthshadow} library \citep{Nir2024_git}, which computes the projected position of the shadow cone at geosynchronous altitude as a function of time. At that distance, the Earth's shadow subtends an angular radius of approximately 8.69°, which we adopt as a fixed shadow radius for all plates. For each of the 645 POSS1-E plates that have at least some area covering north of the celestial equator, we use the plate metadata to reconstruct the shadow trajectory across individual plates over an assumed 50-minute exposure window (i.e., approximately the average over the northern hemisphere plates). Note that the set of in-shadow plates is a property of the POSS1-E survey, and is independent of the SPF datasets. 

Each plate is cropped to a 6$^{\circ}$ × 6$^{\circ}$ square aligned with the cardinal directions. Once the shadow coverage is calculated, SPFs are denoted as: (1) \textit{fully shadowed} — inside the shadow at every time step throughout the entire exposure; (2) \textit{partly shadowed} — inside the shadow at some point during the observation but not the entire time; and (3) \textit{unshadowed} — never within the shadow. Of the 645 plates, five have a fully-shadowed region (plates 728, 08L2, 08L9, 08RS, and 090W); six have only a partly-shadowed region; and the remaining have no shadow interaction at all; one partially shadowed plate (0AVK) crosses the celestial equator. The total number of plates that are at least partially shadowed (11) is consistent with the total (``$\sim 10$'') reported in \cite{villarroel2025aligned}.  Following \citet{villarroel2025aligned}, both the fully shadowed and partly shadowed regions are combined for this analysis and considered to be ``in-shadow''.

The area fractions described above are shown in the top panel of \autoref{fig:fig_shadow_frac}. There are a few notable cases: all SPFs in plate 08L9 lie within the shadow for the full exposure, while there are six plates with hardly any shadow coverage. Because one plate holds most of the SPFs, it dominates the statistical analysis.  \citet{villarroel2025aligned} used $\sim 10$ plates, and claim therefore that a single outlier plate could not be driving their result\textemdash our results suggest the contrary is possible, and we show an example using the case of the related set $W$.

We use both the $W$ and $R$ datasets north of the celestial equator, and we only consider the SPFs that can be confidently assigned to a single plate; this yields 149,301 and 4,782 SPFs for the $W$ and $R$ samples, respectively. Each SPF is then assigned as in-shadow, or not, by testing its sky position against the computed shadow coverage.

\subsection{Statistics}

The expected fraction of SPFs falling in shadowed regions under the null hypothesis of a spatially uniform-random distribution is simply the area-weighted fraction,
\begin{equation}
f_\mathrm{exp} = \frac{\sum_i A_{\mathrm{shadow},i}}{\sum_i \left( A_{\mathrm{shadow},i} + A_{\mathrm{unshadow},i} \right)},
\end{equation}
where the sum includes all 645 plates; this means the fully-shadowed area is $f_\mathrm{exp} = 0.56\%$. The expected number of in-shadow SPFs is then $N_\mathrm{exp} = f_\mathrm{exp} \times N_\mathrm{total}$, where $N_\mathrm{total}$ is the total number of SPFs in the shadowed and unshadowed regions combined. We quantify the departure of the observed count $N_\mathrm{obs}$ from expectation using the signed Poisson significance,
\begin{equation}\label{eq_poisson_sig}
\sigma = \frac{N_\mathrm{obs} - N_\mathrm{exp}}{\sqrt{N_\mathrm{obs} + N_\mathrm{exp}}},
\end{equation}
where negative values indicate a deficit.

For $W$ we find $N_\mathrm{obs} = 649$ against $N_\mathrm{exp} = 729.8$, giving $\sigma = -2.18$. For $R$, we find $N_\mathrm{obs} = 50$ against $N_\mathrm{exp} = 23.1$, giving $\sigma = 3.15$; this is an excess rather than a deficit, driven largely by plate 08L9, which alone accounts for the bulk of the detections in the shadow region (see bottom panels of \autoref{fig:fig_shadow_frac}).

\subsection{Monte Carlo Simulation}

As established earlier in this paper, SPF number density varies significantly from plate to plate and clearly does not follow a Poisson process. Here we show that a naive application of the Poisson significance formula (equation \ref{eq_poisson_sig}) can lead to a misinterpretation of the results. To construct a valid null distribution, we instead use a Monte Carlo procedure. For each of the in-shadow plates, we first construct their shadow templates: the geometry of how the shadow covers the plate; e.g., the template of 08L9 is the full $6^{\circ} \times 6^{\circ}$ square while the template for 0AVK is a tiny sliver. Then we randomly select 11 of the 645 plates, randomly apply the 11 templates, and count the number of SPFs that would fall within the templated area. Summing these counts gives one null trial, from which we compute $\sigma$ using the same formula above. We repeat this procedure $10^5$ times and use the sampled distribution to calculate a $p$-value: i.e., the fraction of trials with $\sigma$ less than the observed value (a one-sided test of the deficit hypothesis).

The resulting distributions are shown in \autoref{fig:fig_bootstrap}. For $W$, we find a deficit of $\sigma = -2.18$; this result sits near the centre of the null distribution ($p = 0.52$), indicating this deficit is not statistically significant: that is, about half of the samples exhibit a larger deficit, and about half exhibit a smaller deficit. For $R$, the observed Poisson significance is $\sigma = 3.15$, which lies in the upper tail of the null distribution ($p = 0.96$), indicating a statistically significant excess.  That is, only 4\% of the samples exhibit an excess as great as $\sigma = 3.15$.

The distributions in \autoref{fig:fig_bootstrap} reveal an important pitfall that arises when applying the Poisson approximation when it is not appropriate (i.e., when the features are not spatially uniform-random). For $W$, the randomly sampled distribution exhibits a wide range of possible Poisson significance values. Our results suggest there is a probability of $\approx 30\%$ to observe at least  $\sigma = -8$ deficit by chance (we note that $\sigma \approx 8$ deficit was estimated by \citet{villarroel2025aligned} using $V$).  This effect is more pronounced in $W$ (compared to $R$) because its larger sample size amplifies the sensitivity to plate-to-plate density inhomogeneity. Thus, there is a greater chance of mistaking a random result to be significant under Poisson statistics with a larger sample size. 
 
Even though we did not use the same dataset as \citet{villarroel2025aligned}, we demonstrate that their analysis is still subject to the effects presented here. Furthermore, \citet{villarroel2025aligned} advocate for large data samples to increase their signal-to-noise ratio. Our results suggest this is a misapplication of Poisson statistics, and that using larger sample sizes amplifies the problem. The claim of a significant deficit of SPFs in Earth's shadow should be reevaluated for $V$ using a Monte Carlo framework for a more reliable signal-to-noise estimate.

For completeness, we ran the same Monte Carlo simulations for a few different scenarios. First, \citet{villarroel2025aligned} ran a test that considered only the central parts of the plates (radius$=2^{\circ}$). Second, we determined whether there was an impact using only the fully-shadowed regions (i.e., excluding the partially-shadowed plates). The results vary, however, they are all consistent with the null hypothesis under the Monte Carlo approach. Therefore, we argue that any of the variations in the shadow calculation just mentioned do not alter our conclusion. It further strengthens our main argument: it is easy for a null result to masquerade as a significant detection using Poisson statistics when the underlying distribution is not spatially uniform-random.

\begin{figure*}[htbp]

    \centering
    \includegraphics[width=\linewidth]{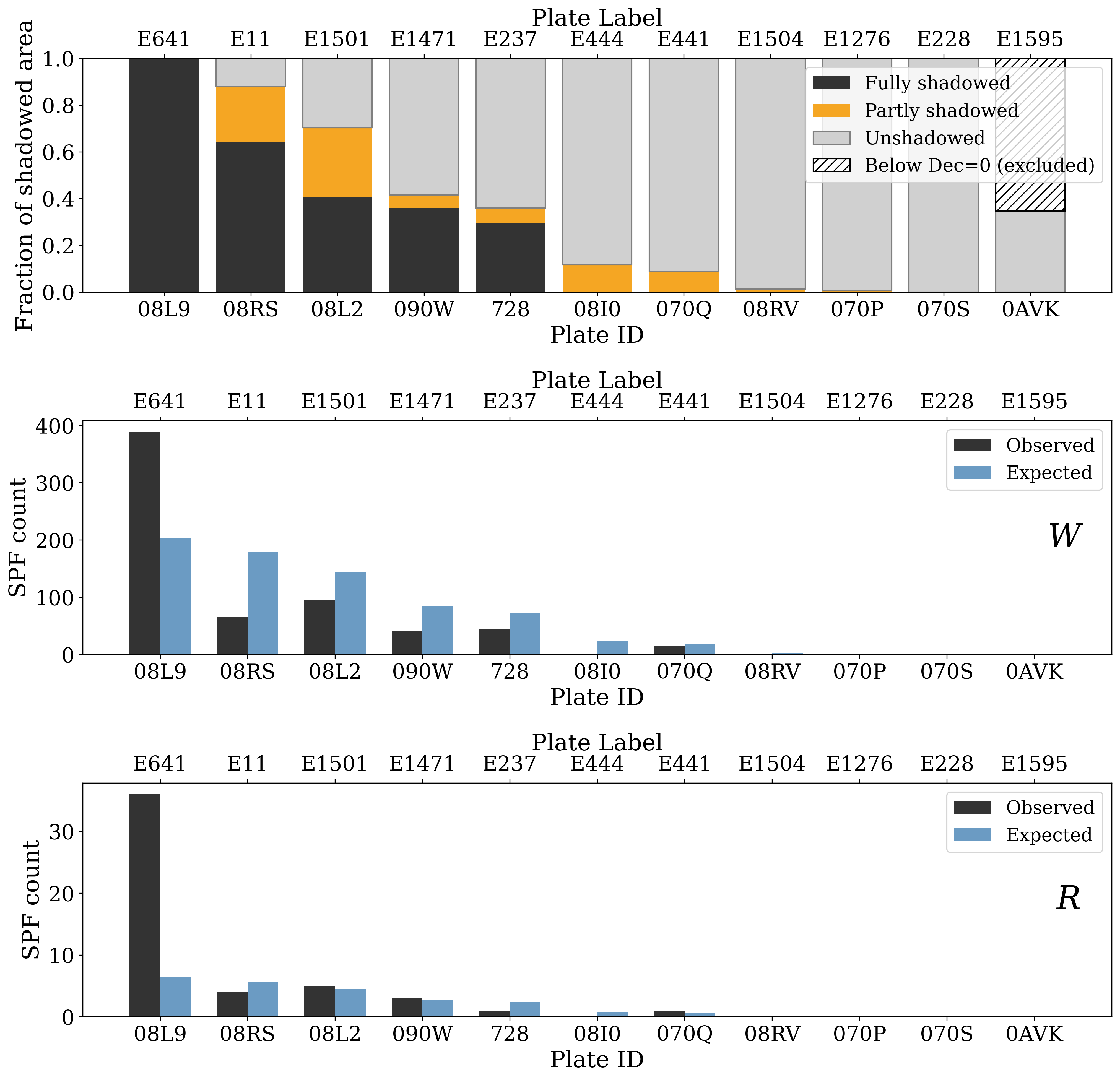}
    \caption{The top bar chart shows the fraction of shadowed area for plates that are at least partly covered by Earth's shadow. Notably, 08L9 is fully shadowed for the full 50 min observation window. The plates 08RV, 070P, 070S, and 0AVK have $<2$\% shadow coverage and, therefore, a marginal effect on the statistics. The number of SPFs that are observed and expected in-shadow for $W$ and $R$ are shown in the middle and bottom panels, respectively.}
    \label{fig:fig_shadow_frac}
\end{figure*}

\begin{figure*}[htbp]

    \centering
    \includegraphics[width=\linewidth]{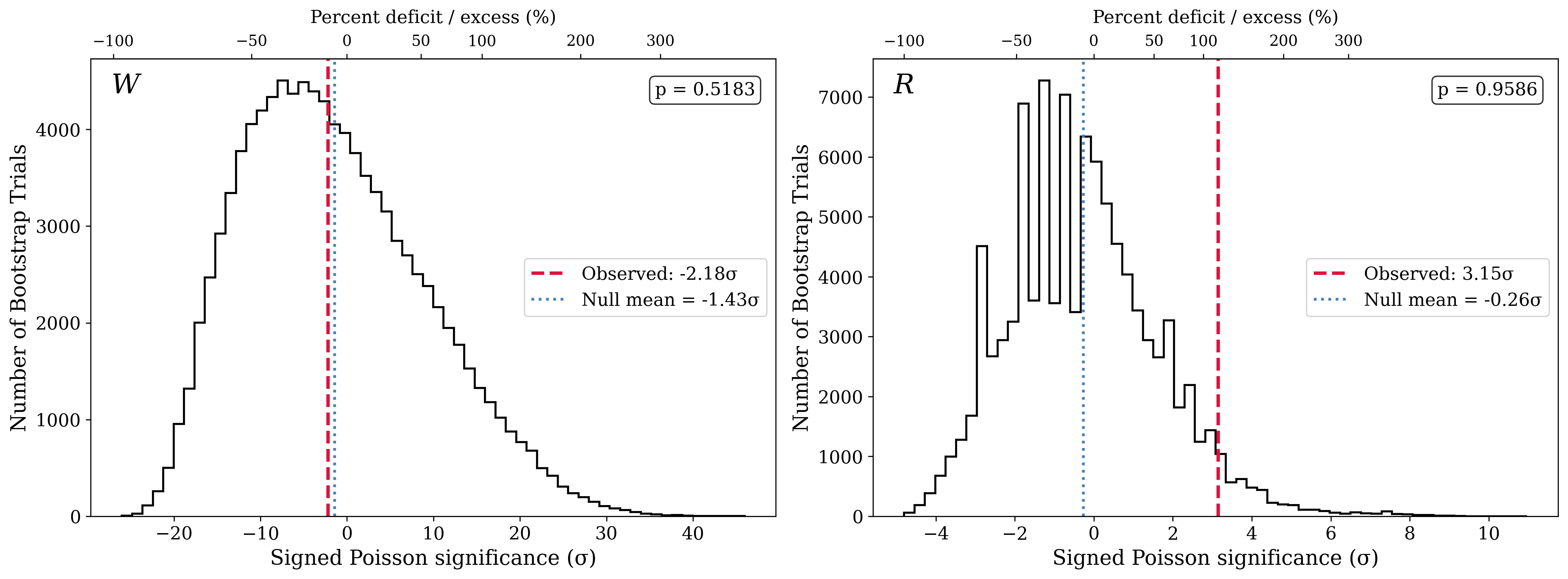}
    \caption{Signed Poisson statistics using the $W$ (left) and $R$ (right) datasets along with their Monte Carlo distributions for $10^{5}$ iterations. Upper axis is the ratio of SPF in-shadow counts: (observed-expected)/expected.  The $p$-value given here is the fraction of trials with $\sigma$ to the left of the observed value (a one-sided test of the deficit hypothesis).  The 
    observed in-shadow deficit for $W$ is not statistically significant, lying very near the null mean ($p = 0.52$): i.e., almost any of the other samples results in a deficit or excess that is larger.  By contrast, the observed in-shadow excess measured for $R$ is moderately significant ($p = 0.96$), since fewer than 4\% of all samples exhibit a larger excess.}
    \label{fig:fig_bootstrap}
\end{figure*}

We argue in many different ways throughout this paper that the spatial distribution of SPFs is not uniform-random.  We do not know the expected distribution of plate- and digitization-related artefacts, which depends on a host of poorly-characterised processes related to the manufacture and degradation of the photographic medium, and the scanning of plates.  A highly nonrandom distribution of these artefacts could significantly affect the relative proportion measured in a tiny fraction of the sky, such as the area covered by Earth's shadow at geosynchronous altitudes. For example, consider the case where the distribution of SPFs is roughly uniform-random except that a handful of plates contains an enormous number of obvious plate artefacts: an example is plate 090R (label E1465) with its 2,149 SPFs in $W$, nearly 20 times the average SPF count per plate.  If the few plates with the overdensity happen to reside outside of the shadow, then the few shadowed plates will contain fewer SPFs per square degree than the population as a whole. As we have seen for the case of set $W$, the in-shadow statistics appear to be dominated by a single plate (i.e., 08L9; see Figure \ref{fig:fig_shadow_frac}).

Another important problem with the in-shadow frequency calculation is that set $V$ was derived from $S$ \citep[][v2]{villarroel2026response} which, as we have seen, was determined by \cite{solano2022discovering} to consist primarily of objects that have not been distinguished from catalogue stars and scan defects. 
At least 95\% (i.e., $1 - N_{\rm R}/N_{\rm V}$) of SPFs in $V$ were discarded by \cite{solano2022discovering} to produce the aggressively-vetted subset $R$ (see Table \ref{tab_data_sets}).  

In summary, in order to fully understand the source of the reported deficit of shadowed SPFs in $V$, it is important to understand the variations in SPF number density at a wide range of scales---from the scale of individual plates as seen in Figures \ref{fig_unusual_patterns} through \ref{fig_offsets_hist2d_x3}, to the scale of the entire sky as seen in Figure \ref{fig_sky_plots_x3}.  Plate degradation and digitization, feature detection algorithms, as well as catalogue filtering and post-processing, all may affect the number density of the shadowed regions while being unrelated to the shadow itself.
For example, it is essential to understand the origin of features like the western hemisphere void that we noted in $R$ (part B of Figure \ref{fig_sky_plots_x3}), and the vertical stripe-shaped void between roughly RA = 90$^\circ$ and RA = 105$^\circ$ (part C of Figure \ref{fig_sky_plots_x3}) that we have noted in $W$ and $R$, and then to examine how shadowed plate footprints and relative SPF counts are affected.

Finally, we call attention to Figure 1 of \cite{villarroel2026response} v2 (upper left), which is a scatter plot of the positions of SPFs in set $V$, to shed light on the origin of the shadow deficit reported in \cite{villarroel2025aligned}.  A comparison of this plot with our Figure \ref{fig_sky_plots_x3}C and its shadowed plate footprints (white boxes), indicates that at least three of the eleven shadowed plates (plate IDs 070P, 08RS, and 090W) overlap the large stripe-shaped void centred on RA $\approx 45^{\circ}$. This includes two of the most significantly shadowed plates (08RS and 090W; see Figure \ref{fig:fig_shadow_frac}).  These two plates are ranked second and fourth overall in terms of the duration and areal extent of shadowing (Figure \ref{fig:fig_shadow_frac}).  This particular stripe-shaped void is not present in $R$ and $W$. This suggests that the incidental overlap of these plates with this stripe-shaped void in $V$, creates a spurious deficit of SPFs in the Earth shadow calculations in \cite{villarroel2025aligned}.  This highlights the importance of characterizing the underlying spatial variation in the data and using an appropriate statistical analysis.

\section{Temporal distribution and alleged correlation with nuclear tests}\label{sec_nuclear}

\begin{figure*}[h!]
\begin{center}
\noindent\includegraphics[width=40pc]{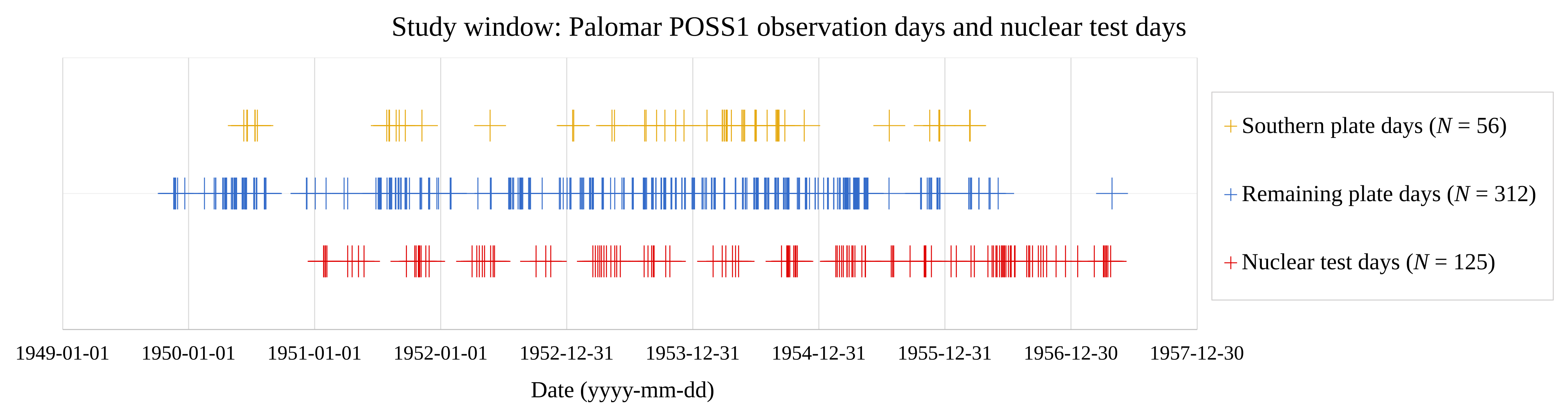}
\caption{From top to bottom, plots of the observation dates (i) with POSS1-E plates solely residing completely in the southern celestial hemisphere; (ii) all remaining plate days (northern celestial hemisphere and/or overlapping the celestial equator); and (iii) the dates of nuclear test explosions.}
\label{fig_plate_nuke_timing}
\end{center}
\end{figure*}

\cite{bruehl2025transients} sought to relate the timing of SPFs in $V$ and the timing of nuclear weapons tests. The time window of the study is defined as spanning November 19, 1949 to April 28, 1957, inclusive, and \cite{bruehl2025transients} used the full 2,718 days for their statistical analyses. However, an examination of the POSS1-E plates metadata reveals that the 937 plate observations were conducted on a total of only 380 days, of which 368 days (906 plates) fall within this study time period. Additionally, we have seen that datasets $V$, $R$, and $W$ have no SPFs from plates fully in the southern hemisphere; this drops the number of relevant observation days down to a maximum of 312.  We show that set $R$ gives very similar results to $V$ when calculating correlation statistics for the dichotomous variables (i.e., SPFs present or not, nuclear tests occurring or not), and that the use of the correct number of relevant observation days (312) dramatically reduces the magnitude and significance of the correlation. Furthermore, as reported in \cite{bruehl2025transients}, ``observed transients'' (SPFs) in $V$ were found on 310 days. That is, SPFs in $V$ were found on plates from almost every relevant observation day (312) during the seven-year POSS1-E survey. Therefore, entirely independent of dataset $R$ (which, incidentally, comprises 289 SPF days), we know that $V$ is highly correlated with the telescope schedule. Finally, and completely independent of any SPF dataset, we find that the telescope schedule is strongly correlated with nuclear tests. We do not address here the continuous variable statistics presented in \cite{bruehl2025transients}.
 We have graphed the dates of the nuclear tests, as well as the POSS1-E plate observation days (``plate days''), in Figure \ref{fig_plate_nuke_timing}.

\subsection{Replication study} 

We start by attempting to replicate the measurement of the correlation between the dichotomous variable (SPFs observed or not) and nuclear tests (nuclear test date $\pm 1$ day), as defined and reported in \cite{bruehl2025transients}. The paper claimed there is a significant (Chi-Square = 6.94, $p = 0.008$) correlation between daily nuclear tests in the study time window and the daily occurrence of SPFs. The left part of Table \ref{tab_nuke_stats} shows the resulting yes/no contingency table presented in Table 1 of \cite{bruehl2025transients} for dataset $V$. We attempted to replicate these results using the three sources  cited in \cite{bruehl2025transients} (see Section \ref{sec_data_sources})
for nuclear test dates, the publicly available SPF dataset $R$, and properties of the publicly available dataset $W$. (The dataset, $V$, used by \cite{bruehl2025transients} was not publicly available at time of writing.) We find a slight count difference in nuclear test dates: \cite{bruehl2025transients} report 124 test dates while we account for 125 dates, likely due to differences in how we parsed the three sources of nuclear test dates. 

Since $R$ (the most vetted dataset available from \cite{solano2022discovering}) and $W$ (the largest dataset available from \cite{solano2022discovering}) do not include plate identification numbers or timestamps, we had to match every detection from these datasets to a POSS1-E plate and its associated timestamp as described in Section \ref{sec_count_variation}.
That is, we excluded SPFs from the datasets which have either no associated plate or ambiguous plate association using this method; this corresponds to about 10\% of the SPFs in $R$ and 12\% in $W$. While there are discrepancies, both in the list of nuclear test dates and in the subset of SPFs that have unambiguous timestamps associated with them, these are trivial and do not change the outcome of the correlation analysis. This is shown in Table \ref{tab_nuke_stats} by the good correspondence of the replication results using dataset $R$ with the same (although incorrect) 2,718 days used by \cite{bruehl2025transients}.

\subsection{Normalization by the true observation window} 

Now that we have successfully replicated the baseline results using dataset $R$, we can estimate the effects of normalizing by just the days on which relevant observations were acquired by the telescope.  SPFs were reported on 310 days out of the 2,718 days in the study period \citep{bruehl2025transients}, of which 255 days occur outside any nuclear testing window and 54 days occur inside a nuclear testing window.  The testing window is defined by \cite{bruehl2025transients} as a day of nuclear testing $\pm1$ day. In order to test for significant correlation between the detection of SPFs and the nuclear testing windows, \cite{bruehl2025transients} compared these numbers (255 days and 54 days, respectively) to the days with and without detected SPFs, that occur within the full time period of 2,718 days. In reality, the POSS1 survey did not record observations every single day out of this time period \citep{minkowski1963national,stsci0000plate}. There are plates available for only 380 days, of which 368 fall within the study period.

In addition, we now use information about datasets $R$, $W$, and $V$ to argue that the 368 days counted within the study period during which plates are available, can be further reduced if we account for the fact that none of these datasets contain SPFs that were sampled from plates fully in the southern celestial hemisphere (see Section~\ref{subsec_data_specific}). Therefore, days when the telescope only observed the southern hemisphere do not count as zero detection days, they count as no-data days, and must be removed from the total count of relevant observation days (Figure \ref{fig_plate_nuke_timing}).

To calculate the number of plate dates to remove due to this overcounting, we use the POSS1-E metadata to see that the footprints of 292 of the 937 POSS1-E plates (31\%) reside entirely in the southern celestial hemisphere without overlapping the celestial equator. However, as explained in Section~\ref{subsec_data_specific}, 
the equatorial plates do contribute a small amount of southern celestial hemisphere SPFs to $W$ (less than 0.1\% ), $R$ (1.4\%), and $V$ (1.4\%) (Section \ref{subsec_data_specific} and Appendix \ref{app_data_sources}).  
Some of the fully southern hemisphere plates were acquired on days when northern hemisphere or equatorial plates were also taken, and these observation days should not be removed from the tally of relevant observation days.
Hence, for the $R$, $W$, and $V$ datasets, this leaves 56 days on which POSS1-E observations were taken fully and exclusively in the southern celestial hemisphere during the study period and are not relevant to this correlation analysis.  Excluding these days, the maximum number of relevant observation days falls from 368 days to 312 days for all three datasets.

Separately, we find 289 unique observation days represented by SPFs in $R$, and 307 in $W$, after matching each of their SPFs to an unambiguous plate; \cite{bruehl2025transients} reports 310 days with SPFs for $V$.  Thus, within the study time window, at least one SPF from $W$ (respectively $R$) dataset was found on 98\% (respectively 93\%) of the days that the POSS1 survey imaged the northern sky. In the case of set $V$, this fraction of relevant POSS1 observation days on which SPFs were detected is 99\%. This suggests that SPFs are not a rare event, they occur on almost every observation day, even within the most filtered set $R$, and their occurrences are almost entirely dictated by the POSS1 observation schedule.   

If we repeat in Table \ref{tab_nuke_stats} the correlation test by replacing the 2,718 days of the full time period with the effective 312 observation days of the POSS1 survey, using dataset $R$, which we have shown in Table \ref{tab_nuke_stats} gives similar results to $V$ when calculating correlation statistics, we see dramatically different conditional probabilities. The $p$-value for a correlation between SPF detection and nuclear testing window goes to $p = 0.1$, which is no longer significant, and the relative risk ratio goes to 1.07 (95\% confidence interval: 1.02 - 1.13). The relative risk ratio here is defined as the conditional probability $\mathbb{P}(\text{SPF detection} | \text{Nuclear test}) = 98.1\%$ divided by the conditional probability $\mathbb{P}(\text{SPF detection} | \text{No nuclear test}) = 91.5\%$. A reasonable interpretation is that SPFs are 7\% more likely to fall within a nuclear test window compared to outside of a nuclear test window (instead of 45\% as claimed in \citep{bruehl2025transients}).  However, even using the correct number of relevant observation days naively estimates the correlation between days with SPFs and days with nuclear tests. We haven't yet considered the underlying baseline correlation between the nuclear tests and the telescope observation schedule itself.

\subsection{Correlation between POSS1 observation schedule and nuclear test schedule} 

Normalizing by the true observation time window (days on which the telescope was actually making observations of the northern celestial hemisphere) is still not satisfactory because it ignores the correlation between nuclear test windows and the observation schedule of POSS1, a correlation independent of plate content. The near-universal occurrence of SPFs on observation days (312), for $V$ (310), $R$ (289), and $W$ (307), ensures that this effect will dominate correlations of SPFs with outside factors. The estimate of this baseline correlation is independent of SPF counts, so we compare the maximum possible relevant observation days (312) with the set of nuclear test windows in the full study period (2,718 days). 

\begin{table*}[h]
    \centering
    \begin{tabular}{|c|c|c||c|c||c|c|}
        \hline\hline
        Source& \multicolumn{2}{c||}{\cite{bruehl2025transients}} & \multicolumn{4}{c|}{This study}\\
        \hline
        Dataset used& \multicolumn{2}{c||}{$V$} & \multicolumn{4}{c|}{$R$}\\
        \hline
        Normalization & \multicolumn{2}{c||}{All days in study} & \multicolumn{2}{c||}{All days in study} & \multicolumn{2}{c|}{ Observation days relevant}\\
         & \multicolumn{2}{c||}{time window(2,718)} & \multicolumn{2}{c||}{time window (2,718)} & \multicolumn{2}{c|}{to the study (312)}\\
        % & \multicolumn{2}{c||}{} & \multicolumn{2}{c||}{} & \multicolumn{2}{c|}{}\\
        \hline
        & \multicolumn{6}{c|}{SPF observed?}\\
        \hline
        Within nuclear testing window? & No & Yes & No & Yes & No & Yes \\
        \hline
        No & 2,116 (89.2\%) & 255 (10.8\%) & 2130 (89.9\%) & 238 (10.1\%) & 22 (8.5\%) & 238 (91.5\%)\\
        \hline
        Yes & 293 (84.4\%) & 54 (15.6\%) & 299 (85.4\%) & 51 (14.6\%) & 1 (1.9\%) & 51 (98.1\%)\\
        \hline\hline
    \end{tabular}
    \vspace{4mm}
    \caption{Left column: reproduction of Table 1 from \cite{bruehl2025transients} for set $V$. Middle column: replication of \cite{bruehl2025transients} using set $R$, included to show that we are able to obtain approximately the same results. Right column: day counts for $R$ obtained after properly normalizing by the true observation window (the correct number of observation days). We have used the most vetted dataset $R$ available from \cite{solano2022discovering} instead of the unpublished set $V$. Due to smaller statistics overall, $R$ contains only 289 days where SPFs were found as opposed to 310 days found by \cite{bruehl2025transients} in the $V$ dataset. The nuclear testing window is defined in \cite{bruehl2025transients} as a day of nuclear testing $\pm1$ day. Percentages in parenthesis show the conditional probabilities of observing (or not observing) an SPF given the presence (or absence) of a nuclear test on that day. Observation days (``plate days'' in Figure~\ref{fig_plate_nuke_timing}) relevant to this study are the days on which at least one plate is sampled  from the northern celestial hemisphere and/or overlapping the celestial equator.
    } 
    \label{tab_nuke_stats}
\end{table*}

Using the same test of correlation between the dichotomous variables that was used for Table \ref{tab_nuke_stats}, we now compare the occurrences of nuclear test windows in the 2,718 days of the full study period with the 312 relevant observation days of the POSS1 survey to estimate the degree of correlation between the observation schedule and the nuclear test windows. We find the relative risk ratio is 1.35 (with a $p$-value of 0.03 and a 95\% confidence interval: 1.03 - 1.78); in other words, a relevant POSS1 observation day is 35\% more likely to fall within a nuclear test window compared to outside of a nuclear test window.

In summary, (i) normalizing by the relevant POSS1 observation days (312) significantly dilutes the significance of the statistical effect between SPF detection and nuclear testing reported in both \cite{bruehl2025transients} and \cite{villarroel2026response} v2; (ii) SPFs are reported by \cite{bruehl2025transients} on at least 99\% of the days on which the POSS1 survey was observing. Out of 312 possible days relevant to these datasets, SPFs are found on 310 days for $V$, 307 days for $W$, and 289 days for $R$, suggesting near-universal occurrence of SPFs on observation days; and (iii) the POSS1 observation schedule has a non-negligible (35\%) likelihood of falling within nuclear test windows compared to outside.  We speculate that the correlation of the telescope schedule and nuclear tests may be due to seasonality, since both the timing of astronomical observations and nuclear testing rely on clear skies and low-wind weather conditions.  

\section{Discussion}\label{sec_discussion}

Research related to SETI and UAP is commonly concerned with searching for unusual signals against a background that is contaminated with artefacts and other sources of noise.  For these fields to advance, it is essential that analyses and conclusions are preceded by rigorous data validation.  This step is essential even in cases where it is extremely difficult.  \cite{solano2022discovering} constructed the dataset $R$ ($N=5,399$) after removing objects that could not be readily distinguished from astronomical objects, plate artefacts, and scan artefacts.  A logical next step is to systematically examine this dataset (or other appropriately filtered dataset) to determine whether any of these features are due to emulsion or other defects and, finally, to determine if any that remain represent true optical sources: that is, whether any originated from a light source in front of the telescope, with a corresponding optical path.  The lack of data validation in \cite{villarroel2025aligned} and \cite{bruehl2025transients} has led to some of the pitfalls that we have described in this paper.

\subsection{Analyzing unvalidated data} 

The underlying problem with the \cite{villarroel2025aligned} and \cite{bruehl2025transients} studies is that the dataset $V$ and its subsets have not been validated in ways that are necessary for the uses to which they have been applied. In reference to the dataset $V$ of ``carefully selected transient samples,'' it is conceded that ``this sample has not been visually inspected. As such, it is expected to contain a substantial number of false positives, including clustered artefacts such as edge fingerprints or other plate defects that contaminate our sample'' \citep{villarroel2025aligned}. The basic question of ``what do the data points represent?'' has not been answered. We have argued that this is absolutely required for any robust scientific result to be derived from these data.  

The necessary data validation for these studies will be challenging to perform. If at some point a candidate SPF for a technosignature is discovered, it will need to be shown that this was created by a true optical path. Section \ref{sec_grbot_litreview} on the search for GRB optical transients revealed both the importance and difficulty of distinguishing between star-like emulsion defects and true optical transients on archival photographic plates. Although most emulsion defects can be identified visually, some need close microscopic examination, and a very few remain indistinguishable from optical flashes. The task is made even more difficult for current researchers using digital scans of such plates. Scans both lose information about the 3-D structures in the emulsion, and introduce new artefacts into the images. Searching for technosignatures in archival plates requires bridging the validation gap between physical plates and digital scans. This involves identifying existing, or creating new, validated optical flashes on physical plates (from known optical sources), tracking them through to digital scans, and doing the same for validated star-like emulsion defects. With these steps, it may be possible to develop a method that can be used to distinguish the two in digital scans and ensure that validated data is used in the analyses. 

A parallel task is to ensure that each optical flash candidate is not related to any known fast astronomical transient or other known sources of optical flashes (see for example \cite{hudec1993optical}). This may benefit from a microscopic examination of known high-altitude satellite glints in photographic plates post-dating POSS1.  In addition, it would be important to study the processes that tend to degrade and contaminate photographic plates, resulting in the creation of new plate artefacts over time.  As previously noted, a morphological analysis with a more stringent filter could be used to remove additional artefacts, in order to highlight a smaller subset that could serve as the focus of these investigations.  

\subsection{Importance of characterizing and addressing background variations}  

If, as in the present case, it is not known what the SPFs represent, the main sources of spatial variation can at least be explored.  This variation should be characterised, modelled, and understood {\it before} examining the relationship to other factors. In the present study, we have highlighted the large-scale heterogeneities in the datasets in question, such as the overdense regions near the Galactic Plane and the stripe-shaped gaps that follow lines of constant right ascension.  We have also displayed the intra-plate distribution of SPFs, finding that SPFs increase in number density toward the plate edges in $W$ and $R$.  We have also shown other noteworthy patterns that occur in plates on an individual basis, including clusters far away from edges (see Figure \ref{fig_unusual_patterns}).  Without showing or modelling the intra-plate distribution of SPFs in $V^{\prime}$ to identify an appropriate remedy, \cite{villarroel2025aligned} removed SPFs located near plate edges (resulting in $V^{\prime\prime}$) and repeated the estimate of the SPF number density in shadowed plates. The result of this analysis was a deficit with only marginal statistical significance.  The key significance of these patterns in number density, which clearly relate to plate geometry, is that they implicate a generative process related to the plates themselves.  This process (e.g., the production of plate emulsion defects) may account for the overwhelming majority of SPFs in $V$, whether in plate interiors or near edges, and which are therefore unlikely to be related to objects in the sky.  

\subsection{Expected spatial distribution of plate artefacts}  

\cite{villarroel2025aligned} uses contradictory assumptions
about the expected spatial distribution of defects, in some places assuming these cluster along edges, and in other places assuming the expected distribution to be ``random'' (e.g., see section 3 on ``Predictions and Expectations'' in \cite{villarroel2025aligned}: ``Plate defects, by contrast, are expected to be randomly shaped and distributed''). Both cannot be true. 

This is prominently revealed by the attention paid to statistical tests of significance to identify correlations, such as a deficit of SPFs in Earth's shadow and SPF alignments. Since we do not understand in detail the processes that can generate plate artefacts, it is not clear that a uniform-random distribution is the correct null hypothesis.  A templating process (see our Section \ref{sec_templating}) could produce plate artefacts exhibiting alignments; for example, the shape created by SPFs in part A of Figure \ref{fig_unusual_patterns} has sharply-defined edges made from dense clusters of SPFs.  A relative deficit of SPFs in a tiny fraction of the sky (e.g., the simulated terrestrial shadow at geosynchronous orbits covered just 1.5\% of the sampled POSS1-E plates) could be entirely explained by excesses of SPFs that may have occurred in plates strongly affected by degradation (see Section \ref{sec_shaddow_freq}), or conversely by deficits arising from features like the large-scale voids visible in Figure \ref{fig_sky_plots_x3} (see Section \ref{sec_sky_dist}).  

Figure 1 in a recent preprint \citep[][v2]{villarroel2026response} includes a plot of SPFs in $V$, which exhibits large-scale spatial heterogeneity similar to what we found for $R$ and $W$, including an additional stripe centred on roughly 45$^{\circ}$.  Our analyses and this figure call into serious question the use of Poisson statistics in \cite{villarroel2025aligned} to quantify the significance of shadowed SPF counts.

\subsection{Inconsistent statements concerning data preparation} 

Ensuring that dataset definitions are clear is essential for others to understand and replicate the reported analyses: it is impossible to build upon work that cannot be readily replicated.  \cite{villarroel2025aligned} contains statements about the extent to which $V$ and its subsets were filtered to remove spurious detections and catalogue objects, which are often inconsistent with each other or inconsistent with descriptions in \cite{solano2022discovering} about the preparation of the parent sets $A$ and $S$. In \cite{villarroel2025aligned}, Section 2 specifies that it ``shall use carefully selected transient samples in Solano et al. (2022), which... have been matched to several modern surveys to remove variable stars, asteroids, and comets.'' 
But $V$ and its subsets are nowhere defined in \cite{solano2022discovering}, which states that asteroids and variable stars were removed as part of the construction of $R$ ($N = 5,399$), only after removing star-proximate SPFs from $S$ (see Table \ref{tab_data_sets}); this leaves a remainder of $N=9,171$, which is clearly too small to be the source of $V$.  The removal of comets is mentioned nowhere in \cite{solano2022discovering}.  
These and other inconsistent statements have made it impossible to accurately reconstruct the definitions of datasets.  The definition of $V$ was almost completely changed in \cite{villarroel2026response} v2, which has claimed that members of $W$ were not specifically removed from $S$ as part of creating $V$.  We have highlighted additional inconsistent remarks about the dataset $V$ and its derivatives in Appendix \ref{app_data_sources}.

\subsection{Circular reasoning}\label{sec_circular}

Circular reasoning involves assuming the conclusion of an argument as part of its premises. 
\cite{villarroel2025aligned} and \cite{bruehl2025transients} make widespread use of the word ``transient'' to describe what we have termed ``SPFs'' (Selected POSS1 Features).  The word ``transient'' implies that the features in question are in fact images of short-lived light sources, before this has been demonstrated.   For example, consider this statement from page 2 of \cite{villarroel2025aligned}:
\begin{itemize}
\item
``Clarifying the origin of these transient events is therefore not only of astrophysical interest but also of potential importance for one of the most enigmatic and consequential questions facing science today.''    
\end{itemize}
This text uses the words ``transient events'' to refer to features on photographic plates, which the same study has elsewhere acknowledged may primarily consist of plate artefacts.  This leading statement implicitly skips a critical unanswered question.  This is akin to asking ``is this feature a UFO or a plane?'' without first establishing that it represents an optical flash that was emitted by or reflected from an object in front of the telescope.

\begin{figure}[h!]
\begin{center}
\noindent\includegraphics[width=20pc]{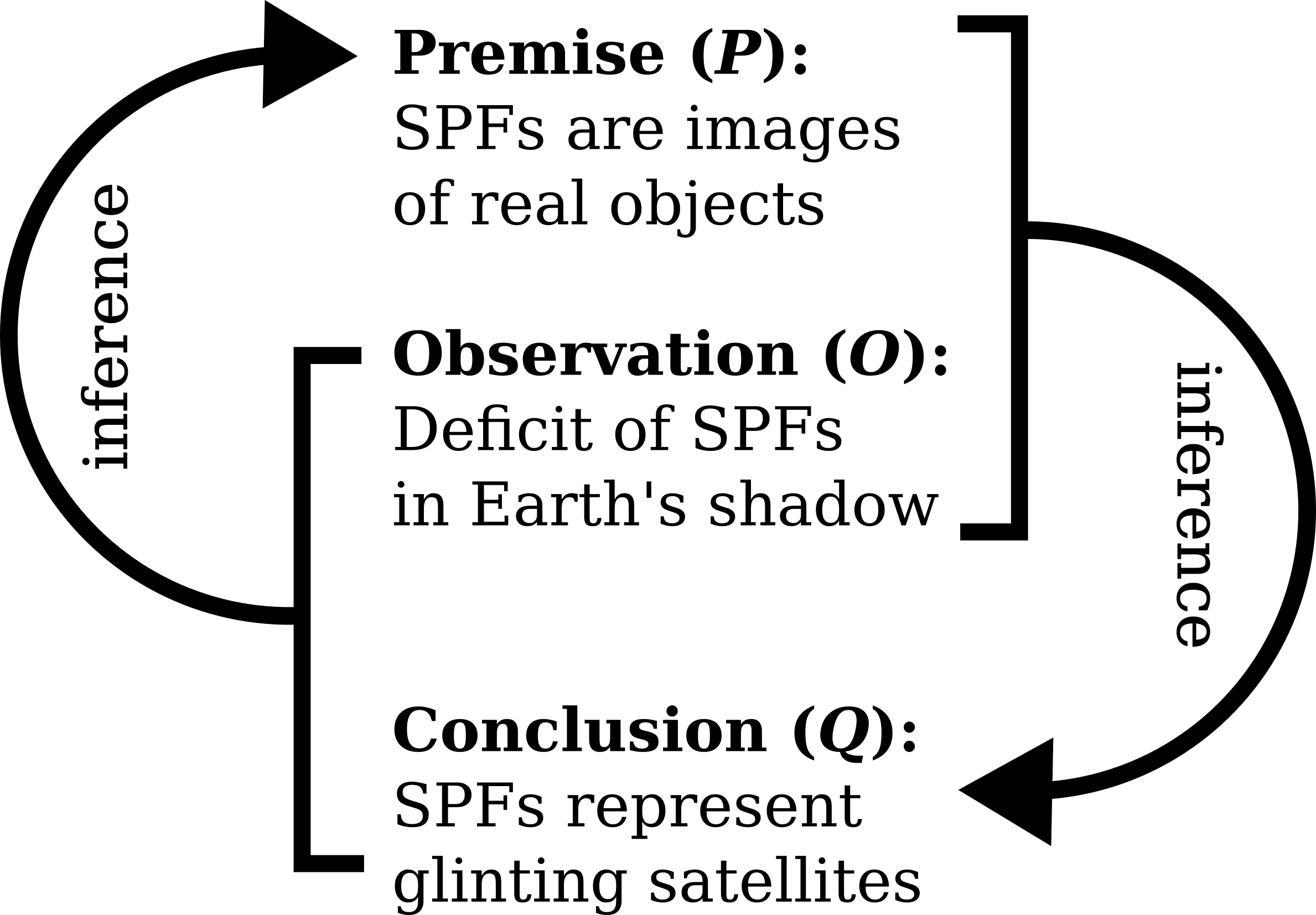}
\caption{Illustration of an example of circular reasoning applied in \cite{villarroel2025aligned}, in which a premise and observation are used to justify a conclusion, while elsewhere the observation and conclusion are also used to justify confidence in the premise.}
\label{fig_circularity}
\end{center}
\end{figure}

Using the {\it result} of an inferential analysis as an independently-demonstrated fact can also lead to circular reasoning. \cite{villarroel2025aligned} invokes specific inferred correlations (deficit in the Earth's shadow, sporadic alignments, correlation with timing of nuclear tests) as evidence for the validity of the data underpinning the analysis.  The circularity of this argument is illustrated in Figure \ref{fig_circularity}.  That is, if the SPFs are valid images and not plate artefacts, then the calculated deficit of SPFs in the shadow and the positive temporal correlation with nuclear tests are leveraged to suggest that these represent images of real objects or events in outer space.  But how do we know that the SPFs are valid images?  The answer from \cite{villarroel2025aligned} appears to be, at least in part: because of a calculated deficit of SPFs in the shadow and a positive temporal correlation with nuclear tests. This circular argument results from an analysis that appears to confirm a hypothesis (specific nonrandom distributions), used to explain the measurements {\it as well as} support their validity.  We give two examples of this device from \cite{villarroel2025aligned}:

\begin{itemize}

\item ``...authentic events may coexist on the same plate as numerous star-like defects, which makes it essential to apply independent diagnostics such as alignment statistics and Earth's shadow tests.'' (pg. 3)

\item ``Even if individual events remain uncertain, Bruehl \& Villarroel (2025) shows statistically significant correlations between subsets of the transient sample in Solano et al. (2022) and historical nuclear activity and aerial anomalies. This alone contradicts the idea that the entire sample consists of plate defects.'' (pg. 3)

\end{itemize}

The first example suggests that if there is a deficit in the shadow, the data should be understood as representing valid measurements (images of objects in front of the telescope) and that these measurements represent glinting objects in orbit around Earth; and if there isn't, then the data is taken to represent neither of these. 

The second example suggests that if there is correlation with nuclear testing, then the data should be understood as representing valid measurements, and that these measurements represent UAP; likewise, the absence of a correlation would be taken to imply neither of these.

To escape the circular reasoning, the underlying validity of the measurements must be established independently of the inferential analysis, which could in turn be cited as support for the conclusion.

As discussed elsewhere in this paper, the ``observations'' that this circular argument relies upon (Figure \ref{fig_circularity}) have been undermined empirically for sets $W$, $R$, and $V$. For example, Sections \ref{sec_spatial_dist} and \ref{sec_shaddow_freq} determined that the assumption of a uniform-random background distribution of SPFs is false, undermining a critical assumption required to draw the statistical inference about an in-shadow SPF deficit.  In Section \ref{sec_nuclear}, the correlation between SPFs and nuclear tests is revealed to be determined largely or entirely by the Palomar telescope observation schedule.

\section{Conclusions}

Searching telescopic observations from the pre-Sputnik era for unidentified luminous and reflecting objects is a novel approach to searching for the existence of technosignatures in proximity to Earth.  It is vital, however, that these searches work from datasets that have been independently validated to represent true optical transients.  In our critique of \cite{villarroel2025aligned} and \cite{bruehl2025transients}, we have pointed out issues with replicability stemming from ambiguous definitions of datasets, as well as issues related to analyses and reported results, many of which arise from a lack of specific and comprehensive validation.  In summary, we offer the following list of conclusions.

\begin{enumerate}

\item {\it Important previous work not addressed}: We conducted a review of optical transient searches in photographic plates, in the context of historical searches for the optical counterparts of Gamma Ray Bursts (GRBs). Following two decades of work, even tentative identification of optical flashes required close microscopic analysis of glass plates, which \cite{villarroel2025aligned} did not perform.  This and preceding work \citep{villarroel2021exploring,villarroel2022glint,solano2022discovering,solano2024bright} have not heeded the primary lessons from this literature.  If robust criteria for identifying optical flashes in archival plates  are eventually identified, this leaves the problem of how to rule out alternative sources of optical flashes such as astronomical fast transients and flashing lights on 
aircraft.

\item {\it Ambiguous dataset definitions}: We have noted discrepancies between the definitions and uses of datasets of Selected POSS1-E Features (SPFs) in \cite{solano2022discovering} and \cite{villarroel2025aligned}.  In searching for SPFs that are not likely to be plate artefacts, digitization artefacts, or catalogue objects, \cite{solano2022discovering} discarded 98.1\% of dataset $S$ ($N = 298,165$), leaving just 5,399 SPFs in $R$. In spite of this, \cite{villarroel2025aligned} analysed the set $S$ and subsets comprising roughly 35\% of $S$ (the subsets $V$ ($N=107,875$) and $V^{\prime}$ ($N=106,339$)); \cite{bruehl2025transients} focused entirely on the set $V$.  Based on \cite{solano2022discovering}, $R$ is the subset of $S$ least likely to contain confounding objects and artefacts, and hence the one most likely to contain genuinely unidentified features. 
However, as mentioned, $R$ itself requires validation by microscopic inspection of individual SPFs \citep{greiner1990discrimination} if a robust set of criteria for identification of optical flashes can be established.  In addition, before any dataset derived from \cite{solano2022discovering} is used for technosignature searches with statistical methods, the sources of noise in these datasets must be researched, understood, and statistically characterised, and precise hypotheses must be defined.

\item {\it Manual evaluation of unidentified SPFs}: We performed a manual review of 10\% of the images of SPFs in set $R$, finding that 4-5\% appear to be obvious artefacts or catalogue stars with high proper motion, despite the extensive filtering in \cite{solano2022discovering}.

\item {\it Small-scale spatial heterogeneity}: The SPFs in subsets of $S$ examined here exhibit distinctive spatial distributions in some plates that are clearly related to the plate boundaries and the celestial coordinate system (Figures \ref{fig_unusual_patterns}, \ref{fig_radial_density}, and \ref{fig_offsets_hist2d_x3}), implicating generative processes that relate to the manufacture, storage, handling, and digitization of photographic plates, environmental conditions, or the subsequent processing of the datasets.  We have discussed the possible templating of SPF placements via zones of contact with other objects and surfaces, or preferential exposure of some regions to degradational processes.  We have also noted that the increase in SPFs with distance from plate centres is qualitatively consistent with the results of experimental studies of endemic plate emulsion defects \citep{greiner1987search}.

\item {\it Large-scale spatial heterogeneity}: The on-sky distribution of SPFs in $W$ reveals windows in right ascension in which there is a marked deficit of SPFs not present in the POSS1-derived $M$ dataset of likely celestial objects.  We have also highlighted a void above the Galactic plane in the western celestial hemisphere in set $R$.  The scatter plot of SPF locations in the unreleased dataset $V$ (see Figure 1 of \cite{villarroel2026response} v2) shows that similar, very pronounced spatial heterogeneity is also a feature of that dataset.    

\item
{\it No statistically significant deficit in the shadow}: We simulated the terrestrial shadow on the northern hemisphere POSS1-E plates. We found 11 shadowed plates in total, five of which contain a region that is shadowed for the entire exposure duration.  We estimated the deviation (deficit or excess) of SPF counts in the 11 shadowed plates with respect to the unshadowed plates, using sets $R$ and $W$.  To understand whether this result is statistically exceptional, we projected the computed shadow patterns onto a different set of 11 randomly-selected plates and repeated this procedure $10^5$ times. We found no statistically significant deficit of in-shadow SPF counts for $R$ and $W$.  We found a very broad null distribution of deviations, consistent with a highly heterogeneous spatial distribution of SPFs.  Separately, we note that two of the five most heavily-shadowed plates overlap a large, stripe-shaped void in $V$, which is not in $R$ or $W$ (see Figure 1 of \cite{villarroel2026response} v2). This may largely account for the in-shadow deficit reported in \cite{villarroel2025aligned}. 

\item 
{\it Inappropriate use of Poisson statistics}: The aforementioned findings about the spatial distribution of SPFs, which depart strongly from uniform-random, show that the use of Poisson estimates of statistical power are not appropriate for the shadow deficit analysis for datasets $R$, $W$, and $V$.  

\item {\it Incorrect normalization for nuclear test correlation study}: Properly normalizing by the number of relevant observation days reduces the magnitude and significance of the correlation between SPFs and nuclear tests as reported in \cite{bruehl2025transients} and \cite{villarroel2026response} v2.  We note that SPFs in $V$ are found on 310 of the 312 observation days, suggesting that the residual correlation may be explained by the statistically significant incidental correlation between nuclear tests and the Palomar telescope observation schedule during the POSS1 survey. This correlation may, in turn, depend on factors that influence both of these activities, such as seasonal effects.  

\item {\it Circular reasoning}: We have also highlighted examples of circular reasoning. Of special note is an argument invoking the result of a contingent inferential analysis (i.e., a deficit of SPFs in the Earth's shadow) to provide justification for confidence in the validity of measurements (i.e., that the SPFs are images of real objects in front of the telescope) as well as the conclusion (i.e., that these images represent glinting objects in outer space).

\end{enumerate}

In summary, our study suggests that before proceeding further with the research program described in \cite{villarroel2025aligned} and \cite{bruehl2025transients}, far more work is needed to clearly define, validate, characterise, and understand the datasets under scrutiny.  First, clear and precise descriptions of how the datasets were constructed are essential for replication and understanding of the work.  Second, SPFs should be independently validated as real images of objects in front of the telescope. This will require the development of methods for confidently distinguishing optical flashes from artefacts, which previous experiments, motivated by the search for GRB counterparts, were not able to establish (see Section \ref{sec_grbot_litreview}).  Systematic examination of artefacts and true flashes, as well as an understanding of how the complex morphology of SPF images are captured by the digitization process, will be essential for this vital step.  Third, the spatial distribution of SPFs should be characterised in detail and the sources of variation must be understood; this is essential for selecting appropriate assumptions about the null distribution ahead of any statistical analyses to examine correlations with other factors. 

The research program addressed in \cite{villarroel2025aligned} and \cite{bruehl2025transients} is pursuing an intriguing question. The effort so far has been instrumental in motivating the contemporary search for technosignatures near Earth using specially-designed instrumentation, such as telescope arrays for triangulating the positions and characterizing the kinematics of anomalous interplanetary objects \citep{villarroel2024searches, villarroel2025cost}.  Although in some ways more complicated owing to the abundance of human spacecraft in orbit around Earth, this direction will avoid many of the challenges and pitfalls plaguing searches in historical photographic plates, and potentially enable the kind of rapid progress made by GRB optical counterpart research after 1996.

\vspace{5mm}

\section*{Acknowledgements}

KK thanks Tony Gorman for his support of University at Albany Project X (UAPx).  We are grateful to Doug Buettner, Tejin Cai, Mike Cifone, Nigel Hambly, Eric Keto, and Matthew Szydagis for helpful comments.  We would like to acknowledge two anonymous reviewers for very helpful comments and suggestions.

Sets $R$ and $W$ are based on data from the SVO archive of vanishing objects in POSS I red images. The archive was developed in the framework of the Spanish Virtual Observatory (https://svo.cab.inta-csic.es) project funded by MCIN/AEI/10.13039/501100011033/ through grant PID2020-112949GB-I00. The system is maintained by the Data Archive Unit of the CAB (CSIC -INTA).

ChatGPT 5 was used for assistance with coding some routines in Python and bash used in the analysis.  AI was not used in the writing of the paper text, or to generate ideas that informed this work, or as an integral part of any analysis (i.e., no machine learning models were used).

\appendix

\section{Ambiguities in dataset definitions}\label{app_data_sources}

As previously mentioned, the origins and reported sizes of $V$ (used in \cite{bruehl2025transients}) and its subsets $V^{\prime}$ and $V^{\prime\prime}$, used for key calculations in \cite{villarroel2025aligned}, are inconsistent in \cite{villarroel2025aligned}.  We raise this issue because this makes replication extremely difficult.  We have listed some of these inconsistent statements here:

\begin{itemize}

\item  From \cite{villarroel2025aligned}, pg 15:  ``We use the transient candidates from \cite{solano2022discovering}, but with the additional requirement that they have no counterparts within $5^{\prime\prime}$ in Gaia, Pan-STARRS and NeoWise.'' The initial candidate selection pool described in \cite{solano2022discovering}, which we have called $S$, already did not include SPFs near Gaia or Pan-STARRS objects \citep{solano2022discovering}. Removing NeoWISE-proximate SPFs from $S$ is consistent with our definition of set $P$ in Table \ref{tab_data_sets} (i.e., $S-W$ or $A-(a_0 \cup W)$), which has an expected size of $N=126,412$.  As previously mentioned, \cite{villarroel2026response} v2 has since clarified that $V$ overlaps with $W$.

\item From \cite{villarroel2025aligned}, pg 15: ``we restrict our analysis to objects in the northern hemisphere (decl. $> 0^{\circ}$). This yields a sample of 106,339 transients, which we use for our study'' and later in the same section: ``in our actual transient data set, only 349/107,875...'' where 107,875 is clearly the total size of the set.  We have taken this latter estimate to be the size of $V$ and the former to be size of $V^{\prime}$.  These differ by 1.4\%, consistent with the proportion of southern hemisphere SPFs in $R$.

\item From \cite{villarroel2025aligned}, pg 3: ``we shall use carefully selected transient samples in Solano et al. (2022), which average 167 transients per plate'' and later ``The transient sample is based on 635 unique photographic plates'', where $167 \times 635 = 106,045 \approx 106,399$.

\item From \cite{bruehl2025transients}, pg 6: ``initial transient dataset consisted of a list of 107,875 transients''.  

\item  From \cite{villarroel2025aligned}, pg 15: ``We apply the published methodology and statistical framework to a published sample of POSS-I transients from Solano et al. (2022)''. As discussed, the only datasets that have been published from \cite{solano2022discovering} are $R$ and $W$. The sets $S$, $V$, and $V^{\prime}$ used in \cite{villarroel2025aligned} and \cite{bruehl2025transients} had not been published at time of writing. 

\end{itemize}

\begin{figure}[h!]
\begin{center}
\noindent\includegraphics[width=23pc]{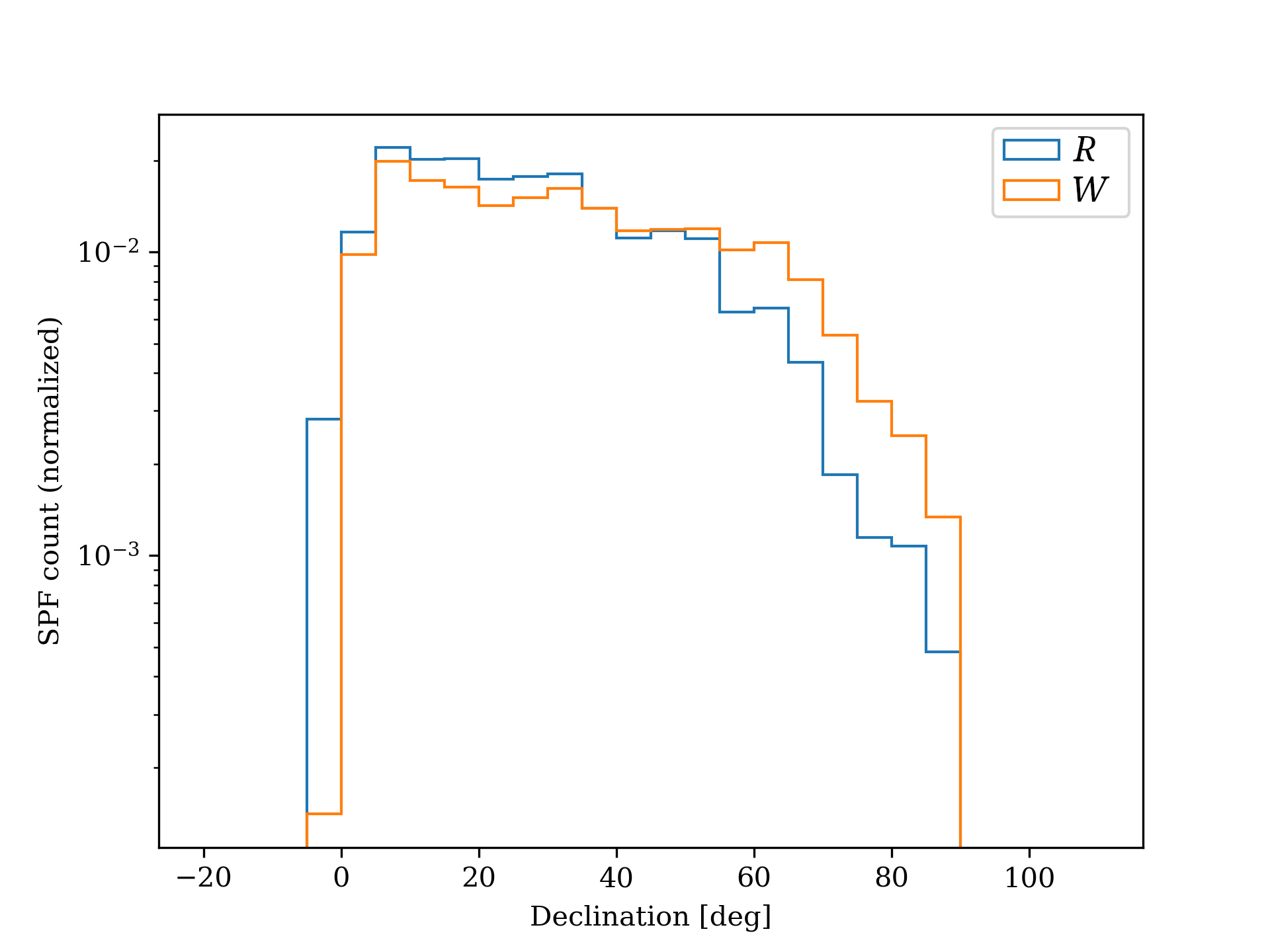}
\caption{Histogram of the SPF counts as a function of declination in datasets $R$ and $W$, suggesting that SPFs in the parent dataset ($S$) and its other derivatives (e.g., $V$) were overwhelmingly sampled from plates whose footprints reside entirely within or that overlap the northern celestial hemisphere.  This was important for (i) understanding how the parent set was sampled, since this hemispherical bias is not mentioned in \cite{solano2022discovering}; (ii) verifying that we have correctly inferred the size of set $V$ (i.e., it has the same proportion of SPFs in the southern celestial hemisphere as $R$), and (iii) for identifying all dates on which observations were acquired, which is vital for assessing the correlation between SPF observations and nuclear tests in Section \ref{sec_nuclear}. The decrease in SPFs approaching the north pole is an effect of geometry (i.e., where there is less area per degree of declination).}
\label{fig_declination_histogram}
\end{center}
\end{figure}

Although set $V$ is attributed to \cite{solano2022discovering}, none of these definitions and reported sizes are consistent with the definitions and sizes of datasets defined in that earlier study. \cite{solano2022discovering} also did not report removing southern hemisphere SPFs or sampling in the southern hemisphere.  The histogram of SPF counts as a function of declination for sets $R$ and $W$ is shown in appendix Figure \ref{fig_declination_histogram}, illustrating a deficit of sampling from the southern hemisphere plates.  The proportion of SPFs in the southern celestial hemisphere is 1.4\% in the case of set $R$ and 0.07\% in the case of set $W$.  Based on the reported sizes inferred for $V$ ($N=107,875$) and $V^{\prime}$ ($N = 106,399$), the proportion of southern-hemisphere SPFs in $V$ is 1.4\%, consistent with the proportion in $R$.

\bibliography{bv-rebuttal}{}
\bibliographystyle{paslike}

\end{document}